\documentclass[10pt]{iopart}
\usepackage{iopams}  
\expandafter\let\csname equation*\endcsname\relax
\expandafter\let\csname endequation*\endcsname\relax
\usepackage{amsmath, amsthm, amssymb}
\usepackage{graphicx} 
\usepackage[leftcaption]{./sidecap}
\usepackage{epsfig}
\usepackage{dcolumn}    % Align table columns on decimal point
\usepackage{xspace} % Sensible space treatment at end of simple macros

\newcommand{\vek}[1]{\boldsymbol{#1}}

\begin{document}

\title[]{Time-domain inspiral templates for spinning compact binaries in quasi-circular orbits
described by their orbital angular momenta 
}

\author{A Gupta$^{1}$ \& A Gopakumar$^{1}$}
\address{$^1$ Department of Astronomy and Astrophysics, Tata Institute of Fundamental Research, Mumbai 400005,
India}
\ead{arg@tifr.res.in, gopu@tifr.res.in}

\date{\today}

\begin{abstract}

We present a prescription to compute the time-domain gravitational wave (GW) 
polarization states associated with 
 spinning compact binaries inspiraling along quasi-circular orbits.
We invoke 
the orbital angular momentum $\vek L$
rather than its Newtonian counterpart $\vek L_{\rm N}$ to describe 
the binary orbits while the two spin vectors are freely specified 
in an inertial frame associated with the initial direction of the total
angular momentum.
We show that the use of $\vek L$ to describe the orbits leads
to additional 
1.5PN order amplitude contributions to the two GW polarization states
compared to the $\vek L_{\rm N}$-based approach and 
discuss few implications 
of our approach.
Further, we provide
a plausible prescription for GW phasing based on certain theoretical considerations and 
which may be treated as the natural circular 
limit to GW phasing for spinning compact binaries 
in inspiraling eccentric orbits 
[Gopakumar A and Sch{\"a}fer G 2011 {\em Phys. Rev. D} {\bf 84} 124007].

\end{abstract}

% 04.30.-w      Gravitational waves
% 04.25.Nx      Post-Newtonian approximation; perturbation theory; related approximations
% 97.60.Lf      Black holes
% 98.35.Jk      Galactic center, bar, circumnuclear matter, and bulge (including black hole and distance measurements)

\submitto{\CQG}

% \maketitle

\section{Introduction}

  Gravitational waves from coalescing compact binaries containing at least one
spinning component are promising GW sources for 
the second-generation laser interferometric detectors like
advanced LIGO, Virgo and KAGRA \cite{Harry10,FA09,KS11}. 
The detection of GWs from such binaries and  the subsequent 
source characterization crucially depend on accurately modeling
temporally evolving GW polarization states, $h_{+}(t)$ and $h_{\times}(t)$, from 
such binaries during their inspiral phase
 \cite{CFPS93}.
Fortunately, inspiral phase of the compact binaries 
can be accurately described by the post-Newtonian (PN) approximation to general relativity \cite{LB_lr}.
At present,  $h_{+}(t)$ and $h_{\times}(t)$  associated with non-spinning compact binaries 
inspiraling along quasi-circular orbits have GW phase evolution accurate to 3.5PN order 
and  amplitude corrections
%that are 3PN accurate \cite{BFIJ,BDFI,BFIS_08}. Recall that the 3.5PN and 3PN orders correspond to 
that are 3PN accurate \cite{BFIJ,BDFI,BF3PN}. Recall that the 3.5PN and 3PN orders correspond to 
corrections that are accurate to relative orders 
$(v/c)^7$ and $(v/c)^6$ beyond the `Newtonian' estimates, where $v$ and $c$ are the orbital 
and light speeds, respectively.

 Obviously, GWs from compact binaries containing Kerr BHs should be extracted from the noisy 
interferometric data  by employing temporally evolving  $h_{+}(t)$ and $h_{\times}(t)$
that incorporate spin effects very accurately and the dominant spin effect arises  due to
the general relativistic
spin-orbit coupling \cite{LK_95}. 
In the PN terminology, the
spin-orbit coupling enters the orbital dynamics formally at 1PN order and for 
moderate spin values $ v_{\text {spin}} \ll c$,
 the coupling numerically appears at the $ {\cal O}(1/c^4)$ level or at the 2PN order \cite{DS_88}. 
However, for  $ v_{\text {spin}} \sim c$, the spin-orbit contributions manifest at slightly more dominant 
 $ {\cal O}(1/c^3)$ level or the 1.5PN order \cite{LK_95}.
We define the spin of a compact object as
$ \vek S = G\, m_{\text {co}}^2\, \chi\, {\vek s}/c$, where
$m_{\text {co}}, \chi$ and $\vek s$ are its mass, Kerr parameter and 
a unit vector along $ \vek S$, respectively.
This definition implies that for a maximally spinning Kerr BH $\chi=1$ and 
for neutron stars  typical upper limit to $\chi$ values should be $\sim 0.4$ \cite{eos}.  
At present, all contributions to the GW phase evolution are available to the next-to-leading order (2.5PN order),
while amplitude corrected $h_{+}(t)$ and $h_{\times}(t)$ are computed to 2PN order for 
spinning compact binaries in quasi-circular orbits \cite{LK_95,FBB,BBF,ABFO,BFH}.
% In comparison, the ready-to-use amplitude corrected GW polarization states are only available to
%2PN order for binaries in quasi-circular orbits \cite{ABFO,BFH,GG1}
Further, it is customary to employ 
the Newtonian orbital angular momentum $\vek L_{\rm N} = \mu\, \vek r \times \vek v $,
where $\mu, \vek r$ and $\vek v$ are the reduced mass, orbital separation and 
velocity, respectively, to specify these quasi-circular orbits.  

   In this paper we provide a prescription to generate the time-domain 
amplitude corrected  $h_{+}(t)$ and $h_{\times}(t)$ for spinning compact binaries
inspiraling along quasi-circular orbits, described by
their orbital angular 
momenta. Note that the amplitude corrected  $h_{+}(t)$ and $h_{\times}(t)$ refer to
GW polarization states that are PN-accurate both in its amplitude and phase.
We begin by describing our GW phasing approach while considering
spinning compact binaries influenced by the 
leading order general relativistic spin-orbit coupling and the
2PN accurate inspiral dynamics (following the literature, we term accurate modeling of 
temporally evolving GW polarization states as `GW phasing').
We discuss, in detail, certain implications of our approach 
where we freely specify the two spins 
at the initial epoch in an inertial frame, 
defined by the initial direction of the total angular momentum $\vek j_0$.
In contrast, it is customary to invoke a $\vek L_{\rm N}$-based orbital triad 
to specify the two spins at the initial
epoch in the literature \cite{LK_95,ABFO,BCV} . 
Our attempt to invoke $\vek L$ while constructing inspiral templates 
%with the help of $\vek L$ is
is motivated by the following considerations.
We observe that a seminal paper that explored the inspiral
dynamics of spinning compact binaries and the influences of precessional 
dynamics on $h_{+, \times}(t)$ employed $\vek L$ to describe their binary 
orbits  \cite{ACST}. 
Moreover, the $\vek L$ variable naturally appears while 
invoking  the canonical formalism of Arnowitt, Deser, and Misner
to describe the dynamics of compact spinning binaries \cite{ADM,SSH}.
We also discuss the theoretical consequence of 
employing the precessional equation appropriate for $\vek L$ 
during the numerical evolution of $\vek L_{\rm N}$.
We show that it can lead to certain anomalous 3PN order terms in the
differential equation for the orbital (and GW) phase 
evolution.
The fact that we invoke $\vek L$ to describe the binary orbits 
leads to additional 1.5PN order amplitude corrections to
$h_{+}(t)$ and $h_{\times}(t)$ compared to
equations~(A2) and (A3) in ~\cite{ABFO} that provide 
amplitude corrected GW polarization states while 
employing  $\vek L_{\rm N}$ to describe the binary orbits.
We provide the reason for the presence of these additional terms 
in our approach and obtain explicit expressions for them.
This implies that these contributions along with 
equations~(A2) and (A3) in ~\cite{ABFO} would lead to 
the fully 1.5PN accurate amplitude corrected $h_{+}(t)$ and $h_{\times}(t)$
in our approach, provided their $\iota$ and $\alpha$ variables specify 
the orientation of $\vek L$ rather than $\vek L_{\rm N}$ (see figure 1 in \cite{ABFO}).
%The above statement also requires that the angular variables $\iota$ and $\alpha$
%that appear in equations~ (A2) and (A3) of  ~\cite{ABFO} 
%provide the orientation of $\vek L$ rather than $\vek L_N$.
%represent $\vek k$ rather than $\vek l$.

Let us emphasize that 
it is equally valid to invoke either $\vek L$ or $\vek L_{\rm N}$
to describe orbits associated with spinning compact binaries.
This implies that the equations~(A2) and (A3) in ~\cite{ABFO}
indeed provide the fully 1.5PN accurate amplitude corrected $h_{+}(t)$ and 
$h_{\times}(t)$ for spinning compact binaries in quasi-circular orbits,
described by $\vek L_{\rm N}$.
% while invoking $\vek L_{\rm N}$.
It will be desirable to employ the appropriate differential equation for $\vek L_{\rm N}$, namely
our equation~(\ref{Eq_lNdot}),  in order to make sure that there are no 
anomalous 3PN order terms in the differential equation for the orbital phase
evolution.
% These 3PN order anomalous terms should be clearly avoided 
% while constructing inspiral templates for spinning compact
% binaries that incorporate higher order spin effects.
We note that the recent detailed PN computations should allow one,
in principle, to write down a PN-accurate differential equation for the orbital 
phase while incorporating the 
3.5PN order 
next-to-next leading order spin-orbit interactions 
and 4PN order spin-spin interactions \cite{MBFB_2012,HS11, HSS13}.
%http://adsabs.harvard.edu/abs/2011AnP...523..783H
%http://adsabs.harvard.edu/abs/2013AnP...525..359H
However, our GW phase evolution is identical to what is provided 
in  ~\cite{ABFO} as both these investigations incorporate only the dominant 
1.5PN accurate spin-orbit effects.
% {\it It should be noted that the GW evolutions in this paper and in \cite{ABFO}
% are identical as we both have considered only 1.5PN order spin-orbit contributions
% to the differential equation for phase evolution. 
% }

  Subsequently, we introduce a plausible prescription to do GW phasing influenced by the
following few theoretical considerations.
These include our observation that  $\omega_{\rm orb}$
naturally exists in a  $\vek L_{\rm N}$-based orbital triad such that 
%usually defined by the relation
 $  \omega_{\rm orb} \equiv  
 \dot {\vek n} \cdot \vek \lambda $, 
where $\vek \lambda = \vek l \times \vek n$ and $\dot {\vek n} = d \vek n/dt $:
$\vek n$ and $\vek l$ are unit vectors along $\vek r $ and $\vek L_{\rm N}$, respectively.
Further, it turns out that 
%We are also influenced by the observation that 
$ \int \omega_{\rm orb}(t') dt'$ (and 
its multiples) can provide GW phase evolutions for spinning binaries 
only when orbital inclinations are tiny as evident from equations~(3.16) 
and (3.17) in ~\cite{ABFO}.
Therefore, we provide a prescription that allows us to impose 
the effects of gravitational radiation reaction on
the conservative evolution of various angles present in the 
expressions for $h_{+}(t)$ and $h_{\times}(t)$ without invoking 
$\omega_{\rm orb}$. 
This approach may be treated as the natural circular limit of GW phasing
for compact binaries in inspiraling eccentric orbits, detailed in \cite{GS11},
as this prescription also requires the orbital energy as the PN expansion parameter.
The above approach looks similar to Taylor-Et approximant, 
introduced in ~\cite{Et} for non-precessing binaries, which turned 
out to be undesirable for data analysis purposes involving 
GWs from non-spinning compact binaries as detailed in \cite{BIPS}.
Further investigations will be required to probe the physical implications of 
our theoretical arguments and it should not be treated as an alternative to
the orbital-like frequency $\omega_{\rm orb}$ based approach. At present, our
aim is to point out the theoretical subtleties involved in the phasing.

   The paper is organized as follows. In section~\ref{abfo_way}, we provide details 
of performing GW phasing for spinning compact 
binaries described by $\vek L$, list features of 
our approach and probe the 
consequence of employing precessional equation of $\vek L$ while numerically evolving 
$\vek L_{\rm N}$. An approach  that does not require 
$\omega_{\rm orb}$ while computing 
the time-domain  $h_{+}(t)$ and $h_{\times}(t)$ for spinning compact binaries
%, associated with a comoving triad linked to $\vek l$,
is  detailed in  section~\ref{GS_way}, while 
section~\ref{Sec_dis_con} provides our conclusions and possible extensions.

\begin{figure}[!ht]
\begin{center}
\includegraphics[width=85mm,height=88mm]{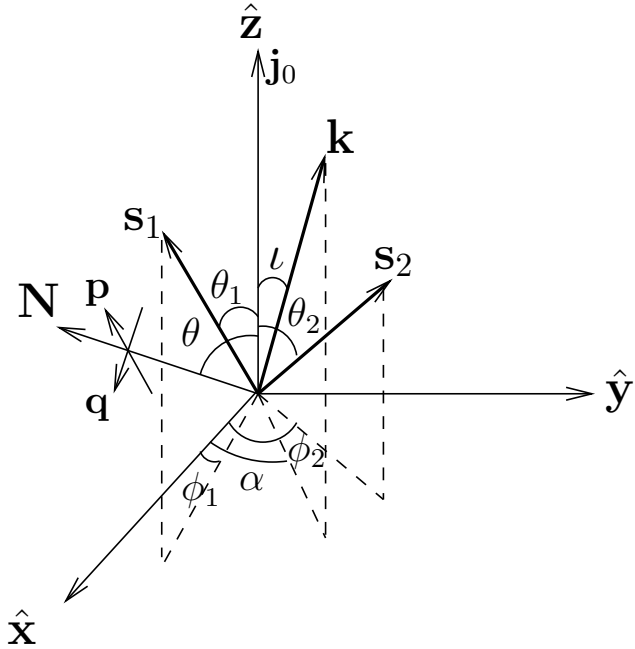}
\end{center}
\caption{  The Cartesian coordinate system where 
the $\vek {\hat z}$ axis points along 
$\vek j_0$, the direction of total angular momentum at the initial epoch.
We display angles that characterize the orbital and spin angular momentum vectors, denoted 
by $\vek k, \vek s_1$ and $\vek s_2$, while the line of sight vector $\vek N$ is 
in the $x-z$ plane. The dashed lines depict projections of various vectors on the $x-y$ plane.
This inertial Cartesian coordinate system is also known as the source frame in the literature.
}
\label{figure:frame}
\end{figure}

 \section{ GW phasing for spinning binaries 
in quasi-circular orbits specified by $\vek L$ and an orbital-like frequency} 
\label{abfo_way}

 We begin by listing formulae required to obtain amplitude corrected GW polarization 
states from the transverse--traceless (TT) part of the radiation field,
$h_{ij}^\text{TT}$:
\begin{subequations}
\label{eq:definition_hp_hx}
\begin{align}
h_{+} &= \frac{1}{2} \left(p_i p_j - q_i q_j \right) h_{ij}^\text{TT}
\,, \\
h_{\times} &= \frac{1}{2} \left(p_i q_j + p_j q_i \right) h_{ij}^\text{TT}
\,,
\end{align}
\end{subequations}
where the orthogonal unit vectors  $\vek p $ and $\vek q $ live
in a plane transverse to the line--of--sight unit vector
$\vek{N}$ defined as $\vek{N} = \vek{R'}/R'$, where 
 $R' = |\vek{R'}|$ is the radial distance from the
observer to the binary (see figure~\ref{figure:frame}).
Following ~\cite{ABFO}, the vectors  $\vek p $ and $\vek q $
are defined with the help of $\vek N$ and $\vek j_0$, a 
unit vector along the direction of the total 
angular momentum at the initial epoch, as
\begin{subequations}
\label{eq:p_q}
\begin{align}
\vek p &= \frac{\vek N \times \vek j_0}{|\vek N \times \vek j_0|}
\,, \\
\vek q &= \vek N \times \vek p
\,.
\end{align}
\end{subequations}
Amplitude corrected  PN-accurate expressions for  $h_{+}(t)$ and $h_{\times}(t)$ originate from 
$h_{ij}^\text{TT}$ which can be expressed as a Taylor series in terms of $v/c$
for binaries in quasi-circular orbits.
The dominant contribution to $h_{ij}^\text{TT}$ arises from the time varying 
Newtonian order quadrupole moment of the binary and is given by
\begin{align}
\label{eq:definition_h_newton}
h_{km}^\text{TT} \big|_{\text Q}
&= \frac{4 G \mu }{ c^4 R'} {\cal P}_{ijkm}(\vek{N})
\left( v_{ij} - \frac{G m}{r} n_{ij} \right)
\,,
\end{align}
where ${\cal P}_{ijkm}(\vek{N})$ is the transverse traceless
projection operator projecting vectors normal to $\vek{N}$,
$\mu $ being the reduced mass ($\mu=m_1\,m_2/m$) 
while $m$ denotes the total mass, $m = m_1 + m_2 $.
In the above equation, we  denoted the 
components of $\vek{n} = \vek{r}/r$ and  the velocity vector
$\vek{v} = d \vek{r} / dt$ by $n_i$ and $v_i$
and defined $v_{ij} := v_i v_j$, and $n_{ij} := n_i n_j $.
The two resulting  GW polarization states are 
\begin{subequations}
\label{eq:h_plus_and_h_cross_in_n_and_v}
\begin{align}
\label{eq_h+}
h_{+} \big|_{\text Q}
&= \frac{2 G \mu}{c^4 R'}
\bigg\{ 
(\vek{p} \cdot \vek{v} )^2 - (\vek{q} \cdot \vek{v} )^2
-  \frac{G m}{r} \left[ ( \vek{p} \cdot \vek{n} )^2
- ( \vek{q} \cdot \vek{n} )^2 \right] 
\bigg\}
\,, \\
\label{eq_hx}
h_{\times} \big|_{\text Q}
&= \frac{4 G \mu}{c^4 R'}
\left[ (\vek{p} \cdot \vek{v}) (\vek{q} \cdot \vek{v}) - \frac{G m}{r}
(\vek{p} \cdot \vek{n}) (\vek{q} \cdot \vek{n}) \right]
\,.
\end{align}
\end{subequations}
Clearly, these expressions are  for compact binaries in general orbits
and their circular versions are obtained by using $ v^2 = G\, m/r $,
after evaluating above dot products. The expressions for $h_{+}(t)$ and $h_{\times}(t)$
suitable for constructing inspiral templates for spinning compact binaries
inspiraling along quasi-circular orbits can be obtained 
with the help of following few steps.
In the first step, we express $\vek r$ and $\vek v$  in the inertial frame 
$\left ( \hat {\vek x}, \hat {\vek y}, \hat {\vek z} \right )$, where $\hat {\vek z}$ 
points along $\vek j_0$,
with the help of the three Eulerian angles $\Phi, \alpha$ and  $\iota$
defined in the inertial frame. The resulting expression for 
$\vek r$ reads
\begin{subequations}
%\label{eq:r_and_v}
\begin{align}
\label{eq:r_xyz}
\vek r &= r\vek n \,, \mbox {where } \\
%\end{align}
%where 
%\begin{align}
\label{eq:n_xyz}
\vek n &= (-\sin \alpha\,\cos \Phi-\cos \iota\,\cos \alpha\,\sin 
\Phi,\,   
 \cos \alpha\,\cos \Phi   \nonumber \\
&\quad-\cos \iota\,\sin \alpha\,\sin \Phi\, ,\,\sin
\iota\,\sin \Phi)\,,   
\end{align}
\end{subequations}
and the velocity vector becomes 
$ \vek v =  d\vek r/dt = r \, d {\vek n}(\Phi, \iota, \alpha)/dt $
for circular orbits
 \cite{KG05}.
For our purpose it is
convenient to represent various vectors present in 
equations~(\ref{eq:h_plus_and_h_cross_in_n_and_v}) in the 
comoving frame defined by the 
triad $( {\vek n},{\vek {\xi}} = \vek k \times \vek n,{\vek k} )$, where 
$\vek k$ is the unit vector along $\vek L$.
This is easily achieved with the help of three rotations involving the 
three Eulerian angles appearing in the expression for $\vek r $ \cite{DS_88, KG05}.
The components of $\vek r, \vek v, \vek p, \vek q $ and $\vek N$ in the comoving frame
are given by
\begin{subequations}
\label{eq:p_q_r_v}
\begin{align}
\label{eq:r_comov}
\vek r &= r\vek n\,,  \\  \label{eq:v_comonv}
\vek v &=  
%\dot r\, \vek n +
r\, \biggl (  \frac{ d \Phi}{dt} +   \frac{ d \alpha}{dt}\,
\cos \iota
 \biggr ) \, \vek \xi
+ r\, \biggl (  \frac{ d \iota }{dt}\, \sin  \Phi 
- \sin \iota \, \cos \Phi\,  \frac{ d \alpha}{dt} \biggr )\, \vek k
\,, \\
\vek p &= (\sin \Phi \, \cos \iota \, \sin \alpha - \cos \Phi \, \cos \alpha)\, \vek n 
+ (\sin \Phi\, \cos \alpha +  \cos \Phi\, \cos \iota\, \sin \alpha)\, \vek \xi \nonumber \\
&\quad-\sin \iota\, \sin \alpha\, \vek k \,, \\
\vek q &= (-\sin \Phi \, \sin \iota \, \sin \theta  - \cos \Phi\, \sin \alpha \, \cos \theta  
- \sin \Phi \, \cos \iota\, \cos \alpha \, \cos \theta )\, \vek n  \nonumber \\
&\quad+(-\cos \Phi \, \cos \iota\, \cos \alpha \, \cos \theta  - \cos \Phi \, \sin \iota \, \sin \theta   
+ \sin \Phi \, \sin \alpha \, \cos \theta )\, \vek \xi  \nonumber \\
&\quad+(\sin \iota \, \cos \alpha \, \cos \theta - \cos \iota \, \sin \theta)\, \vek k \,,  \\
\vek N &=(-\cos \Phi\, \sin \alpha\, \sin \theta - \sin \Phi \, \cos \iota\, \cos \alpha\, \sin \theta       
+ \sin \Phi \sin \iota \cos \theta)\, \vek n  \nonumber \\
&\quad+ (\sin \Phi\, \sin \alpha \, \sin \theta             
- \cos \Phi\, \cos \iota \, \cos \alpha \, \sin \theta + \cos \Phi \, \sin \iota \cos \theta )\, \vek \xi        \nonumber \\
&\quad+ (\sin \iota \, \cos \alpha \, \sin \theta + \cos \iota \, \cos \theta)\, \vek k \,,
\end{align}
\end{subequations}
where we used equations~(\ref{eq:p_q}) for $\vek p$ and $\vek q $ and let $\vek j_0 = (0,0,1)$ 
and $\vek N = (\sin \theta, 0, \cos \theta)$
in the inertial frame as shown in figure~\ref{figure:frame}.
The vectors defining the comoving frame, namely $\vek n, \vek \xi$ and $\vek k$, 
have following components in the 
inertial frame 
\begin{subequations}
\begin{align}
\vek \xi &= (\sin \alpha\,\sin \Phi - \cos \iota\,\cos \alpha\,\cos \Phi\, ,\, 
 -\cos \alpha\,\sin \Phi  \nonumber \\
& \quad-\cos \iota\,\sin \alpha\,\cos
  \Phi\, ,\,\sin \iota\,\cos \Phi)\,,  \\
\vek k &= (\sin \iota\,\cos \alpha\, ,\,\sin \iota\,\sin
\alpha\, ,\,\cos \iota)
\,,
\end{align}
\end{subequations}
while  $\vek n$ is specified by equation~(\ref{eq:n_xyz}) such that $\Phi$ measures the orbital phase 
from the direction of ascending node in the $x-y$ plane. 

  It is now fairly straightforward to obtain explicit expressions for 
$h_{+, \times}\big|_{\text{Q}}$ in terms of various angular variables 
with the help of equations~(\ref{eq:p_q_r_v}) and the resulting expressions read
\begin{subequations}
\label{eq:hplus_hcross}
\begin{align}
h_+|_{\rm Q} &=  \frac{2\, G\, \mu\, v^2}{c^4\, R'}\,\Biggl\{\left(\frac{3}{2}\,\cos^2 \iota
- \frac{3}{2}\right)\, (1-C_{\theta}^2) \cos 2\Phi  \nonumber \\
&\quad - \left(1+\cos \iota\right)\,S_{\theta}\,C_{\theta}\, \sin \iota\, \cos (2\Phi+\alpha)  \nonumber \\
&\quad - \frac{1}{4}( \cos^2 \iota+ 2\,\cos \iota 
+1 )\,(1+C_{\theta}^2)\, \cos (2\alpha + 2\Phi)  \nonumber \\
&\quad - \frac{1}{4}( \cos^2 \iota- 2\,\cos \iota 
+1 )\,(1+C_{\theta}^2)\, \cos (2\alpha - 2\Phi)  \nonumber \\
&\quad-S_{\theta}\, C_{\theta}\, \sin \iota \, \cos \iota \, \cos (\alpha-2 \Phi)    \nonumber \\
&\quad+S_{\theta}\, C_{\theta}\, \sin \iota \cos (\alpha-2\Phi)\Biggr\} \,,     \\
%%%%%%%%%%%%%%%%%%%%%%%%%%%%%%%%%%%%%%%%%%%%%%%%%%%%%%%%%%%%%%%%%%%%%%%%%%%%%%%%%%%%%%%%%%%%%%%%
h_{\times}|_{\rm Q} &= \frac{2\, G\, \mu\, v^2}{c^4\, R'}\,\Biggl\{(1
-\cos \iota)\, S_{\theta}\,  \sin \iota\, \sin (\alpha- 2 \Phi)   \nonumber \\
&\quad - (1+\cos \iota )\,S_{\theta}\, \sin \iota \, \sin (\alpha + 2\Phi)  \nonumber \\
&\quad -\frac{1}{2}(1+ 2 \cos \iota+ \cos^2 \iota)\, C_{\theta} \sin (2\alpha +2\Phi)  \nonumber \\
&\quad -\frac{1}{2}(1- 2 \cos \iota+ \cos^2 \iota)\, C_{\theta} \sin (2\alpha -2\Phi)  
\Biggr\} \,,
\end{align}
\end{subequations}
 where $ v^2/c^2 = ( G\, m\, \dot \Phi/c^3)^{2/3}$ while $S_{\theta}$ and $C_{\theta}$ stand for $\sin \theta$ and $\cos \theta$, respectively. 
To obtain above the expressions from equation~(\ref{eq_h+}) and (\ref{eq_hx}),
we used the Newtonian accurate relation, 
$ v^2=r^2\, \dot \Phi^2 = G\, m/r$ arising from equation~(\ref{eq:v_comonv}) for $\vek v$
and let $(r\, \dot \Phi)/c = ( G\, m\, \dot \Phi/c^3)^{1/3}$. 
%\newline
%{\bf READ CAREFULLY \& let me know}
%\newline
   
    GW phasing for inspiraling binaries containing spinning compact objects 
is performed by first prescribing 
differential equations that provide precessional (conservative) evolution
for various Eulerian angles present in the expressions for $h_{+, \times}(t)$.
Thereafter, one imposes the effect of radiation damping on these
conservative evolutions. We begin by prescribing the differential equation for $\Phi$
based on the following considerations \cite{LK_95}.
With the help of equation~(\ref{eq:v_comonv}) for $\vek v$, we write down the following expression 
for $v^2$ to the desired 1.5 PN order
\begin{equation}
\label{Eq_v2}
v^2 =  r^2\, \dot \Phi^2 + 2\,r^2\, \dot \Phi\, \dot \alpha \, \cos \iota \,,
\end{equation} 
and this leads to $ v = r\, (\dot  \Phi + \cos \iota \, \dot \alpha ) $.
Invoking an orbital-like frequency $\omega_{\rm orb} \equiv v/r $, we write
\begin{equation}
 \dot \Phi = \omega_{\rm orb} - \cos \iota \, \dot \alpha \,.
\end{equation}
It is convenient to introduce a dimensionless parameter
 $x \equiv ( G\, m\, \omega_{\rm orb} /c^3)^{2/3}$,
allowing us to obtain following equation for the conservative evolution of $\Phi$:
\begin{align}
\label{Eq_phidot}
 \dot \Phi =  \frac{ x^{3/2} }{ ( G\, m/c^3)} - \cos \iota\, \dot \alpha \,.
\end{align}
The conservative evolution of $\alpha $ and $\iota$ may be extracted from 
the precessional equations for $\vek k$ as $\alpha $ and $\iota$  
specify $\vek k$ in the inertial frame. The structure of $\dot {\vek k}$
demands us to solve (numerically) precessional equation for $\vek k$
together with those for 
$\vek s_1$ and $\vek s_2$.
The relevant equations, extractable from ~\cite{LK_95, BBF}, read
\begin{subequations}
\label{Eq_dxy_ks1s2}
\begin{eqnarray}
%\begin{align}
\label{Eq_kdot}
{\dot {\vek k}} &=&  \frac{c^3}{Gm}\, x^3\,\Bigg\{ \delta_1\,q\, \chi_1\, 
\left( \vek s_1\times\vek k\right) \label{eq:kdot} 
+\frac{\delta_2}{q}\, \chi_2\, \left( \vek s_2\times\vek k\right) \Bigg\} \,, \\
{\dot {\vek s}_{1}} &=&  \frac{c^3}{Gm}\, x^{5/2}\,\delta_1 \left(\vek k\times \vek s_1\right) \,, \\
{\dot {\vek s}_{2}} &=&  \frac{c^3}{Gm}\, x^{5/2}\,\delta_2 \left(\vek k\times \vek s_2\right) \,,
%\end{align}
\end{eqnarray}
\end{subequations} 
where
$q=m_1/m_2$ and  the Kerr parameters 
$\chi_{\rm n}$ specify the spin vectors  by $ \vek S_{\rm n} = G\, m_{\rm n}^2\, \chi_{\rm n}\,
\vek s_{\rm n}/c $, where the subscript $n$ can take values $1$ or $2$. 
The symmetric mass ratio $\eta = \mu/m$ is required to define 
the quantities $\delta_1$ and $ \delta_2$ as
\begin{subequations}
\begin{align}
\delta_{1} &
%= 2 \eta \left( 1 + \frac{3 m_2}{4 m_1} \right)
= \frac{\eta}{2} + \frac{3}{4} \left(1 - \sqrt{1 - 4 \eta} \right)
\,,
\\
\delta_{2} &
%= 2 \eta \left( 1 + \frac{3 m_1}{4 m_2} \right)
= \frac{\eta}{2} + \frac{3}{4} \left(1 + \sqrt{1 - 4 \eta} \right)
\,.
\end{align}
\end{subequations}
We note that the equations~(\ref{Eq_dxy_ks1s2}) provide precessional dynamics due
to the dominant order spin-orbit coupling.
Further, for the purpose of numerical integration we specify the components of 
$\vek s_1$ and $\vek s_2$ in the inertial frame associated with $\vek j_0$ as
\begin{subequations}
\label{eq:s1_s2}
\begin{align}
\vek s_1 &= \left ( \sin \theta_1\,\cos \phi_1, \sin \theta_1\,\sin \phi_1, \cos \theta_1 \right )\,,\\
\vek s_2 &= \left ( \sin \theta_2\,\cos \phi_2, \sin \theta_2\,\sin \phi_2, \cos \theta_2 \right )\,,
\end{align}
\end{subequations}
as displayed in  figure~\ref{figure:frame}.

    We are now in a position to incorporate the effect of gravitational radiation damping
and this is achieved by 
imposing secular evolution of $x$ in  equations~(\ref{Eq_phidot}) and (\ref{Eq_dxy_ks1s2}).
% Eq_phidot, equations~(\cite{Eq_dxy_ks1s2})
%specifying secular variations in $x$ in the above 
%differential equations that provide conservative temporal evolutions of 
%the relevant Eulerian angles. 
In this paper, we employ the 2PN-accurate expression for $ \dot x $
that requires the usual energy balance
argument and available in \cite{BBF}:
%
%For the present discussion we employ
%following 
%2PN-accurate expression for $ \dot x $ arising from the usual energy balance
%argument, given in ~\cite{BBF}:
%
\begin{eqnarray}
%\begin{align}
\label{Eq_dxdt2}
\frac{ d x}{dt} &=& \frac{64}{5}\frac{c^3}{Gm}\eta\, {x}^5 
\biggl \{
1+x \left [ -\frac{743}{336}-\frac{11\eta}{4}\right ] +4\pi x^{3/2} \nonumber \\
&&+ \frac{x^{3/2}}{12}\,\Biggl [ (-188\,X_1+75\sqrt{1-4\eta})\,X_1\, \chi_1\, (\vek s_1\cdot \vek k) \nonumber \\
&&+(-188\,X_2-75\sqrt{1-4\eta})\,X_2\,\chi_2\, (\vek s_2 \cdot \vek k)\Biggr ]
\nonumber \\
&&+  x^2 \,\biggl [ \frac{34103}{18144}+\frac{13661}{2016}\eta +\frac{59}{18}\eta^2 \biggr ]
\biggr \} \,,
%\end{align}
\end{eqnarray}
where $X_1=m_1/m$ and $X_2=m_2/m$.

  We propose to
solve together equations~(\ref{Eq_phidot}),~(\ref{Eq_dxy_ks1s2}) and 
(\ref{Eq_dxdt2}) numerically,
while invoking the Cartesian components of the precessional equations
and specifying the spin vectors in the inertial frame defined by $\vek j_0$.
Naturally,
  we need to specify initial conditions 
for the Cartesian components of
$\vek k, \vek s_1$ and $\vek s_2$ in the inertial frame.  These components for $\vek s_1$
and $\vek s_2$ are obtained with the help of equations~(\ref{eq:s1_s2}) by specifying all possible
initial values for
$(  \theta_1,\phi_1)$ and $( \theta_2, \phi_2)$.
 The fact that 
we align the total angular momentum
along $z$-axis at the
initial epoch  allows us to equate  the $x$ and $y$ components 
of $\vek J= \vek L + \vek S_1 + \vek S_2$ 
at the initial instant to zero.
This results in the following 
expressions to estimate 
initial  values of  
the Cartesian components of $\vek k$ 
%The   initial estimates for the $k_{\rm x}$, $k_{\rm y}$ and $k_{\rm z}$ read 
\begin{subequations}
\label{Eq_ialpha_ini}
\begin{eqnarray}
%\begin{align} 
k_{\rm x,0}&=&-\frac{G\, m^2}{c\,L_{\rm 2PN}(x_0)}\,
\{X_1^2\, \chi_1\,\sin \theta_{10}\,\cos \phi_{10}  \nonumber \\
&&+X_2^2\, \chi_2\,\sin \theta_{20}\,\cos \phi_{20}\} \,, \\
k_{\rm y,0}&=&-\frac{G\, m^2}{c\,L_{\rm 2PN}(x_0)}\,
\{X_1^2\, \chi_1\,\sin \theta_{10}\,\sin \phi_{10}   \nonumber \\
&&+X_2^2\, \chi_2\,\sin \theta_{20}\,\sin \phi_{20}\} \,, 
%k_{z,0}&=&+\sqrt{1-k_{x,0}^2-k_{y,0}^2} \,,
%\end{align}
\end{eqnarray}
\end{subequations}
where $\theta_{10}$, $\theta_{20}$, $\phi_{10}$ and $\phi_{20}$ are 
the initial values of $\theta_1, \theta_2, \phi_1$ and $\phi_2$, respectively
while $L_{\rm 2PN}(x_0)$ denotes the value of 
%$L_{\rm 2PN}$,
 the 2PN accurate orbital angular
momentum at $x =x_0=x (t_0)$. The relevant analytic expression for $L_{\rm 2PN}$ is
given by \cite{BBF}
\begin{eqnarray}
\label{Eq_L2}
\vek L_{\rm 2PN }&=& \frac{G\,m^2\,\eta}{c}\,x^{-1/2}\,
\biggl \{ 1+ x\bigg[\frac{3}{2}+\frac{\eta}{6} \bigg] 
 +x^2\,\bigg[\frac{27}{8}-\frac{19\eta}{8}+\frac{\eta^2}{24}\bigg] \biggr \} \vek k \,,
%\end{align}
\end{eqnarray}
and we would like to note that the above estimates for 
%initial values for
 $k_{\rm x,0}$ and $k_{\rm y,0}$ do not change 
substantially even if we drop PN corrections in $\vek L_{\rm 2PN }$.
The initial values for $\alpha$ and $\iota$ are 
obtained, as expected, by equating the above expressions for
$k_{\rm x,0}$ and $k_{\rm y,0}$ to 
$\sin \iota\,\cos \alpha $ and $\sin \iota\,\sin
\alpha$, respectively, and numerically solving these coupled 
equations. In our numerical runs, we select those 
solutions that provide positive $\iota$ values 
at the initial epoch.
The bounding values for $x$ are
given by $x_0=2.9\times 10^{-4}\, (m'\, \omega_0)^{2/3} $ 
and $x_{\rm f}=1/6 $, where $m'$ is the total mass of the binary in solar units
and let $\omega_0 = 10\, \pi$ Hz as customary for aLIGO
and we choose the initial phase to be zero ($\Phi_0 =0$).
Finally, we note that the values of $\alpha$ and $\iota$
at every step of our numerical runs are obtained from 
the Cartesian components of $\vek k$ by 
$ \alpha = \tan^{-1}(k_{\rm y}/k_{\rm x})$ and $\iota = \cos^{-1} (k_{\rm z})$.

   We are now in a position to plot the temporally evolving $h_{+, \times}\big|_{\text{Q}}(t)$
for inspiraling compact binaries described by equations~(\ref{eq:hplus_hcross}).
%Eqs.~(\ref{eq:hplus_hcross}).
In figures~\ref{figure:q_1_x} and \ref{figure:q_4_x}, 
we plot $\iota(t)$ and $h_{+, \times}\big|_{\text{Q}}(t)$ for few spin configurations
and two mass ratios ($q=1$ and $4$) in the aLIGO frequency window for maximally spinning
BH binaries. The choice of $m = 50 \, M_{\odot}$ allows us to explore 
the combined effects of precessional and reactive orbital dynamics 
in short time windows spanning around $ \sim 10$ seconds
for both mass ratios.
The spin configurations are chosen so that the more massive BH spin orientations 
vary from relatively smaller to larger values from $\vek j_0$.

 Clearly, the plots in figure~\ref{figure:q_1_x} do not show any 
amplitude modulations and additional numerical runs indicate similar behavior 
even for extreme spin configurations. This is due to the fact that 
for equal mass binaries $|\vek L| >> |\vek S|$, where $\vek S=\vek S_1 + \vek S_2$,
throughout the inspiral 
leading to tiny modulations \cite{LK_95, ACST}. 
%We may also attribute 
This leads to slow reactive evolutions of $\iota$ 
in addition to low initial values
for $\iota$, prescribed by equations~(\ref{Eq_ialpha_ini}), for equal mass binaries.
%as the main contributing factors.
The slower variations for $\iota$  may also be attributed to
our observation that   
$ \vek k \cdot \vek S_{\rm eff}$, 
where $\vek S_{\rm eff}= \delta_1\, \vek S_1 + \delta_2\, \vek S_2$, 
remains a constant during the reactive evolution for 
equal mass binaries. We recall that the conservation 
of $ \vek k \cdot \vek S_{\rm eff}$ allowed 
the derivation of certain Keplerian type parametric 
solution to the underlying dynamics in ~\cite{KG05}.

The plots in figure~\ref{figure:q_4_x} depict temporal 
variations in $ \iota,  h_{ \times}\big|_{\text{Q}}$ and 
$ h_{+}\big|_{\text{Q}}$ for unequal mass maximally spinning
BH binaries. We clearly observe amplitude modulations, for spin-configurations B and C,
where the dominant spin
is misaligned from $\vek j_0$ by $60^{\circ}$ and $120^{\circ}$, respectively.
The comparatively higher and faster secular variations in $\iota$ 
at the later stages of inspiral
can be attributed to the fact that the spin angular momentum of the more massive BH 
dominates over the orbital angular momentum during this stage. 
This is consistent with the deduction of ~\cite{ACST} that pointed out faster $\iota$ 
variations 
for spin angular momentum dominated binaries. 
However, the differences in $\iota$ evolutions, 
evident between Spin-B and Spin-C configurations in figure~3,
prompted us to include variations in $ \vek k \cdot \vek S_{\rm eff}$ 
also contributing to the way $\iota$ evolves.
This is because the spin angular momentum starts dominating 
both these binaries at the same epoch during the late inspiral though we observe
different $\iota$ evolutions.
Further, such binaries have relatively larger 
initial values for $\iota$  
compared to their equal mass counterparts 
in addition to faster secular variations in $\iota$. 
These effects clearly contribute to the observed 
 amplitude modulations in 
$h_{+, \times}\big|_{\text{Q}}(t)$.

% \begin{widetext}

\begin{figure}
\end{figure}

\begin{figure}[!h]
%\centering
\begin{center}
\includegraphics[width=100mm,height=80mm]{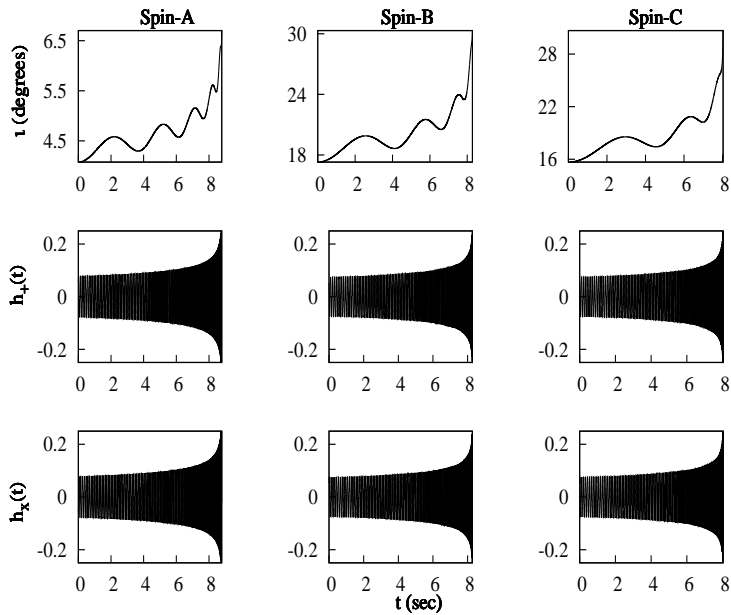}
\end{center}
\caption{Plots showing temporally evolving  
$\iota(t)$, scaled  $h_{+}\big|_{\text{Q}}(t)$ and 
$h_{\times}\big|_{\text{Q}}(t)$ for maximally spinning compact binaries
$(q=1, m = 50M_{\odot})$. The three initial spin configurations 
are denoted by Spin-A:
 \{$\theta_{10}=\pi/6$, $\phi_{10}=\pi/4$, $\theta_{20}=\pi/6$, $\phi_{20}=\pi$\},
Spin-B: \{$\theta_{10}=\pi/3$, $\phi_{10}=\pi/4$, $\theta_{20}=\pi/3$, $\phi_{20}=\pi/2$\}
and  Spin-C: \{$\theta_{10}=2\,\pi/3$, $\phi_{10}=\pi/2$, $\theta_{20}=\pi/4$, $\phi_{20}=\pi/4$\},
while the scaling factor is ($2\, G\, \mu\,/c^2\, R'$).
The theoretically prescribed initial values for $\alpha $ and $\iota$ are
$(-67.5^{\circ}, 4.1^{\circ}), (-112.5^{\circ}, 17.3^{\circ})$ and $(-110.1^{\circ}, 15.7^{\circ}) $, 
respectively. Further, the initial angle between $\vek k$ and $\vek s_1$ for the above three 
spin configurations A, B and C are $31.76^{\circ}$, $76.14^{\circ}$ and $134.52^{\circ}$, respectively
while the initial negative $\alpha$ values, if required, may easily be converted to 
their positive counterparts by subtracting from $2\, \pi$.
}
\label{figure:q_1_x}
\end{figure}

\begin{figure}[!h]
\begin{center}
\includegraphics[width=100mm,height=80mm]{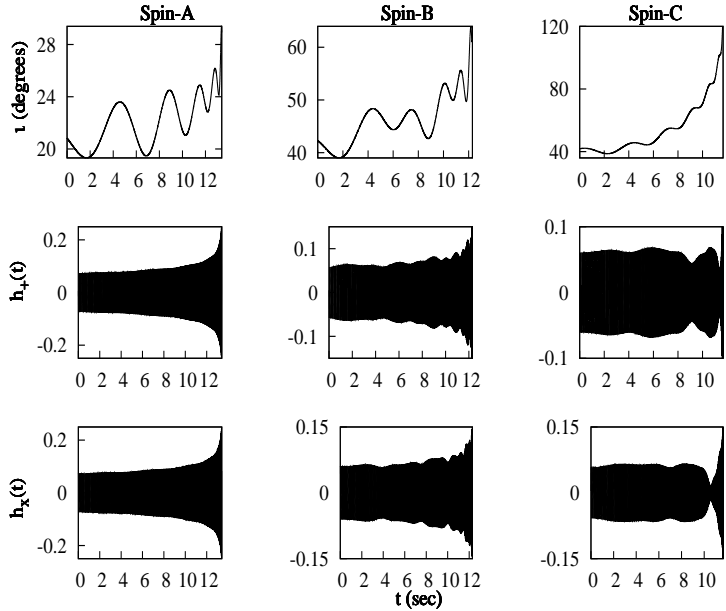}
\end{center}
\caption{ Plots, similar to figure~\ref{figure:q_1_x}, for maximally spinning
unequal mass binaries $(q=4, m = 50M_{\odot})$. The prescribed initial values 
for $\alpha$ and $\iota$, based on equations~(\ref{Eq_ialpha_ini}), for the three initial spin configurations are
$(-132.4^{\circ}, 20.8^{\circ}), (-132.6^{\circ}, 42.3^{\circ})$ and $(-91.9^{\circ}, 41.9^{\circ}) $,
respectively.  The initial angle between $\vek k$ and $\vek s_1$ for the above three 
spin configurations  A, B and C are $50.82^{\circ}$, $102.25^{\circ}$ and $161.80^{\circ}$, respectively.
%and for all plots, the bounding values for $x$ are 
%0.165541, 0.165815 and 0.166176, respectively. 
Spin induced amplitude modulations are clearly visible in
spin configurations that provide substantial evolution for $\iota$.
We observe that $\vek k\cdot \vek S_{\rm eff}$ vary during the inspiral and amplitude of
these variations are initial spin configuration dependent.
}
\label{figure:q_4_x}
\end{figure}

% \end{widetext}

In what follows we probe the consequence of employing 
$\dot { \vek L_{\rm N}} = - (\dot {\vek S_1} + \dot {\vek S_2} )$ 
to numerically evolve $\vek l = \vek L_{\rm N}/| \vek L_{\rm N} |$.
It turns out that it is equivalent to employing an 
orbital averaged expression for $\dot{\vek l}$ \cite{LK_95}.
Invoking an orbital averaged precessional equation for $\vek l$ 
leads to an undesirable feature that the coefficient of
$\vek l$ in the expression for $\dot {\vek n}$ in 
$( {\vek n},{\vek {\lambda}} = \vek l \times \vek n,{\vek l} )$
frame
will not, in general, vanish.
To verify the above observation, let us specify these 
unit vectors in the inertial frame associated with $\vek j_0$
using three angles $\Phi', \alpha'$ and $\iota'$ such that
\begin{subequations}
\label{Eq_n_lam_l}
\begin{align}
\vek n &= (-\sin \alpha'\,\cos \Phi'-\cos \iota'\,\cos \alpha'\,\sin 
\Phi', \label{Eq_n}\,   
 \cos \alpha'\,\cos \Phi'   \nonumber \\
&\quad-\cos \iota'\,\sin \alpha'\,\sin \Phi'\, ,\,\sin
\iota'\,\sin \Phi')\,,   \\
  \vek \lambda &= (\sin \alpha'\,\sin \Phi' - \cos \iota'\,\cos \alpha'\,\cos \Phi'\, ,\, 
 -\cos \alpha'\,\sin \Phi'  \nonumber \\
& \quad-\cos \iota'\,\sin \alpha'\,\cos
  \Phi'\, ,\,\sin \iota'\,\cos \Phi')\,,  \\
\label{Eq_lN}
\vek l &= (\sin \iota'\,\cos \alpha'\, ,\,\sin \iota'\,\sin
\alpha'\, ,\,\cos \iota')\,,
\end{align}
\end{subequations}
where we used {\it primed} variables to distinguish from the Eulerian angles 
present in the $\vek k$-based frame.
It is fairly straightforward to compute the time derivative of $\vek n$ 
and express it in the $( {\vek n},{\vek {\lambda}},{\vek l} )$
frame as
\begin{align}
\label{eq:dndt}
\frac{d {\vek n} }{dt} &=  \biggl ( \frac{ d \Phi'}{dt} + \cos \iota' \,
 \frac{ d \alpha'}{dt} \biggr ) \, \vek \lambda
+ \biggl (  \frac{ d \iota' }{dt}\, \sin \Phi' 
- \sin \iota' \, \cos \Phi'\,  \frac{ d \alpha'}{dt} \biggr )\, \vek l
\,.
\end{align}
The fact that $\vek l = \vek n \times \dot{\vek n} /|\vek n \times \dot {\vek n}|$ 
clearly demands 
that the coefficient of $\vek l$ in the above equation should be zero 
as also noted in ~\cite{ABFO}.
However, if one employs equation~(\ref{eq:kdot}) as the precessional equation for $\vek l$ namely
\begin{align}
\label{Eq_ldot}
 {\dot {\vek l}} &=  \frac{c^3}{Gm}\, x^3\,\Bigg\{ \delta_1\,q\, \chi_1\, 
\left( \vek s_1\times\vek l\right) 
+\frac{\delta_2}{q}\, \chi_2\, \left( \vek s_2\times\vek l\right) \Bigg\} \,, 
\end{align}
then it is possible to show with some straightforward algebra that 
\begin{equation}
\label{Eq_relation}
 \sin \Phi' \, \frac{d\iota'}{dt} - \cos \Phi' \, \sin \iota' \, \frac{d\alpha'}{dt} 
= -\frac{c^3}{G\, m}\, x^3\, \Bigg\{ \delta_1 \, q\, \chi_1\, (\vek s_1 \cdot \vek \lambda ) 
 + \frac{\delta_2}{q}\, \chi_2\, (\vek s_2 \cdot \vek \lambda)\Bigg\} \,.
\end{equation}
It is not very difficult to conclude that 
the right hand side of above expression, in general, is not zero.
The above equation is obtained by employing 
 following equations for $d\iota'/dt $ and $d\alpha'/dt$ 
%To obtain above equation, we employed following equations to obtain  
%the time derivatives of   $\iota'$ and $\alpha'$
\begin{subequations}
\label{Eq_idot_adot}
\begin{align}
 \frac{d\iota'}{dt}&= \frac{c^3}{G\, m}\, x^3\, \Bigg\{  
[\delta_1 \, q\, \chi_1\, (\vek s_1 \cdot \vek n )    
+ \frac{\delta_2}{q}\, \chi_2\, (\vek s_2 \cdot \vek n)]\, \cos \Phi' \nonumber \\
&\quad- [\delta_1 \, q\, \chi_1\, (\vek s_1 \cdot \vek \lambda )
+ \frac{\delta_2}{q}\, \chi_2\, (\vek s_2 \cdot \vek \lambda)] \, \sin \Phi' \Bigg\} \,, \\  
%&=-\frac{c^3}{Gm}\, x^3\,\Bigg\{ \delta_1\,q\, \chi_1\, \sin \theta_1\, \sin(\alpha'-\phi_1) \nonumber \\
%&\quad+\frac{\delta_2}{q}\, \chi_2\, \sin \theta_2\, \sin (\alpha'-\phi_2) \Bigg\} \,, \\
%%%%%%%%%%%%%%%%%%%%%
\frac{d\alpha'}{dt}\,\sin \iota' & =  \frac{c^3}{G\, m}\,x^3\, \Bigg\{ 
 [\delta_1 \, q\, \chi_1\, (\vek s_1 \cdot \vek \lambda )   
+ \frac{\delta_2}{q}\, \chi_2\, (\vek s_2 \cdot \vek \lambda)]\, \cos \Phi'  \nonumber \\
&\quad+[\delta_1 \, q\, \chi_1\, (\vek s_1 \cdot \vek n )
+ \frac{\delta_2}{q}\, \chi_2\, (\vek s_2 \cdot \vek n)]\, \sin \Phi'  \Bigg\} \,,
%&= \frac{c^3}{Gm}\, x^3\,\Bigg\{ \delta_1\,q\, \chi_1\, \Biggl(\cos \theta_1 \, \sin \iota'  \nonumber \\
%&\quad- \sin \theta_1 \, \cos \iota'\, \sin (\alpha'+\phi_1)\Biggr) \nonumber \\
%&\quad+\frac{\delta_2}{q}\, \chi_2\, \Bigg(\cos \theta_2 \, \sin \iota' 
%- \sin \theta_2 \, \cos \iota'\, \sin (\alpha'+\phi_2)\Biggr) \Bigg\} \,,
\end{align}
\end{subequations}
and these equations easily arise from equation~(\ref{Eq_lN}) for $\vek l$ 
and equation~(\ref{Eq_ldot}) for $\dot {\vek l}$.

   It is possible to correct the above inconsistency by using the appropriate
precessional equation for $\vek l$ as noted in \cite{GS11}.
In the covariant Spin-Supplementary-Condition (SSC) pursued here \cite{BO_75},
the precessional equation for $\vek l$ arises from the following PN-accurate 
expression that connects $\vek l$ and $\vek k$ \cite{BBF}
\begin{equation}
\label{Eq_l_k_rel}
 \vek l = \vek k +  \frac{x^{3/2}}{G\, m^2}\, \Biggl\{ \left(-\frac{1}{2}\, S_n
-\frac{1}{2}\,\frac{\delta \, m}{m} \Sigma_n\right)\, \vek n  
+ \left(3S_{\lambda}+\frac{\delta \, m}{m}\, \Sigma_{\lambda}\right)\, \vek \lambda
\Biggr\}  \,,
\end{equation}
where $ \vek {\Sigma} = m \left ( {\vek S_2}/m_2 - {\vek S_1}/m_1\right)$ as
introduced in ~\cite{BBF}, while 
  $\delta m=m_1-m_2$. Further, 
 $ (S_n$, $\Sigma_n)$ and $( S_{\lambda}$, $\Sigma_{\lambda}) $ are the 
components of $\vek S$ and $ \vek {\Sigma}$ along  $\vek n$ and $\vek \lambda$, respectively.
Taking time derivative of the above equation and using
%A straightforward time derivative computation of the above equation using 
equation~(\ref{Eq_kdot}) for $\dot {\vek k}$ gives
\begin{align}
\label{Eq_lNdot}
 \dot {\vek l} &= -2\frac{c^3}{G\, m}\, x^3\, \Bigg\{ \delta_1 \, q\, \chi_1\, (\vek s_1 \cdot \vek n )
+ \frac{\delta_2}{q}\, \chi_2\, (\vek s_2 \cdot \vek n) \Bigg\}\, \vek \lambda \,.
\end{align}
We now compute terms that appear in the coefficient of $\vek l$ in 
the expression for $ \dot {\vek n}$, given by equation~(\ref{eq:dndt}) using
the above equation for $\dot {\vek l}$ and equation~(\ref{Eq_lN}) for $\vek l$.
They are given by 
\begin{align}
 \sin \Phi'\, \frac{d\iota'}{dt}=\cos \Phi'\, \sin \iota' \frac{d\alpha'}{dt}  
&= 2\, \frac{c^3}{G\, m}\, x^3\, \sin \Phi' \, \cos \Phi'\, \Bigg\{ \delta_1 \, q\, \chi_1\, (\vek s_1 \cdot \vek n )  \nonumber \\
&\quad+ \frac{\delta_2}{q}\, \chi_2\, (\vek s_2 \cdot \vek n) \Bigg\} \,.
\end{align}
Therefore, the coefficient of 
$\vek l$ in the expression for $\dot {\vek n}$ computed using the appropriate expression for 
$ \dot {\vek l}$ indeed vanishes as required.

 It turns out that $\dot {\vek n}$ having components along $\vek l$ 
leads to anomalous terms that contribute to $\Phi'$ evolution at the 3PN order
and this is within the consideration of higher order spin effects  
available in the literature. 
 Thanks to very detailed PN computations, it should be possible, in principle, to write 
down a differential equation for $\Phi'$ that incorporates the next-to-next-to-leading order
spin-orbit and spin-spin interactions \cite{MBFB_2012,HS11, HSS13}.
%http://adsabs.harvard.edu/abs/2011AnP...523..783H
%http://adsabs.harvard.edu/abs/2013AnP...525..359H
It should be noted that higher order spin-orbit and spin-spin interactions contribute to
the orbital dynamics at 3.5PN and 4PN orders, respectively, for maximally spinning BH binaries.
%Note that higher order spin effects are known to 3.5PN order while 
%dealing with spin-orbit interactions and to 3PN order while dealing with spin-spin 
%coupling \cite{HSS2010,MBFB_2012}. 
To verify the above statement we observe 
that the definitions $ \vek v = r\, \dot {\vek n}$ and 
$ v = r\, \omega $ imply that $ \omega^2 = \dot {\vek n} \cdot \dot {\vek n}$.
With the help of our equation~(\ref{eq:dndt}) the expression for $ \omega^2$ reads
\begin{align}
 \omega^2 &= \biggl ( \frac{ d \Phi'}{dt} + \cos \iota' \,
 \frac{ d \alpha'}{dt} \biggr )^2 
% \, \vek \lambda
+ \biggl (  \frac{ d \iota' }{dt}\, \sin \Phi' 
%\no
- \sin \iota' \, \cos \Phi'\,  \frac{ d \alpha'}{dt} \biggr )^2\,,
% \vek l\,.
\end{align}

Taking the square root leads to
\begin{align}
 \omega &= \biggl ( \frac{ d \Phi'}{dt} + \cos \iota' \,
 \frac{ d \alpha'}{dt} \biggr ) 
% \, \vek \lambda    
+  \frac{1}{2\, \dot \Phi'} \biggl (  \frac{ d \iota' }{dt}\, \sin \Phi' 
%\no
- \sin \iota' \, \cos \Phi'\,  \frac{ d \alpha'}{dt} \biggr )^2\,,
% \vek l\,.
\end{align}
where $ \dot \Phi'$ obviously stands for $ d \Phi'/{dt}$ and higher order terms 
that are cubic in the time derivatives of $\iota'$ and $\alpha'$ are neglected.
Invoking the fact that $\dot \Phi'$ at the Newtonian order is given by $ x^{3/2}/ (G\,m/c^3)$
and noting that $d \iota'/dt $ and $d \alpha'/dt $ have $c^3\,x^3/( G\,m) $ as the common factor,
we get
\begin{align}
\dot { \Phi'}&= \frac{c^3}{G\,m}\, x^{3/2} \biggl ( 1 + x^{3/2} \, A' + x^{3} \, B'\biggr ) \,,
\end{align}
such that $A'$ and $B'$ are given by
\begin{align}
 A' &= -\frac{\cos \iota'}{\sin \iota'}\Big \{ [\delta_1 \, q\, \chi_1\, (\vek s_1 \cdot \vek \lambda )
+ \frac{\delta_2}{q}\, \chi_2\, (\vek s_2 \cdot \vek \lambda)] \, \cos \Phi'  \nonumber \\
&\quad+ [\delta_1 \, q\, \chi_1\, (\vek s_1 \cdot \vek n )    
+ \frac{\delta_2}{q}\, \chi_2\, (\vek s_2 \cdot \vek n)]\, \sin \Phi'\Big \} \,, \\ 
B' &= -\frac{1}{2}\, \Big \{ \delta_1 \, q\, \chi_1\, (\vek s_1 \cdot \vek \lambda )
+ \frac{\delta_2}{q}\, \chi_2\, (\vek s_2 \cdot \vek \lambda) \Big \}^2 \,.
\end{align}
These expressions mainly arise from the curly brackets present on the RHS of equations~(\ref{Eq_relation}) and (\ref{Eq_idot_adot}).
It should be evident that the $B'$ terms appear at the 3PN order while $A'$
terms enter $\dot{\Phi}$ expression at the 1.5PN order.
It is not very difficult to infer that these  $B'$ terms arise due to the non-vanishing
$\vek l$ component in the expression for $\dot {\vek n}$
and therefore are unphysical in nature.
This computation  shows that the anomalous $\vek l$ components in 
the expression for $\dot {\vek n}$
contribute to $\Phi'$ evolution at an order that is within the consideration
of higher order spin effects available in the literature.
% Note that these 3PN order unphysical contributions to 
% $\dot \Phi'$ should not influence the $\Phi'$ evolution in ~\cite{ABFO}.
% This is because their $\Phi'$ evolution was governed by a 1.5PN accurate
%  differential equation, namely equation~(3.10) in \cite{ABFO},
% that incorporated only the leading order spin-orbit effects.
%It should be noted that these unphysical terms play no role in the investigations of ~\cite{ABFO}
%that probed the leading order spin effects appearing at the 
%1.5PN order in the phase evolution.

Let us note that we do not freely specify the two spins in a non-inertial orbital 
triad as usually done in the literature~\cite{BCV,ABFO}.
Recall that we freely specify the two spin vectors in an inertial source
 frame associated with $\vek j_0$, namely the usual source frame, at the initial epoch.
This choice allowed us to estimate the initial $x$ and $y$
components of $\vek k$ in the source frame 
uniquely in terms of the orientations of 
$\vek s_1, \vek s_2$ and other intrinsic binary parameters at the initial epoch, as given by
equations~(\ref{Eq_ialpha_ini}).
This is mainly because  at the initial epoch one is allowed to let
$\vek j_0$ along the $z$-axis without any loss of generality and thereby
allowing us to impose that the 
$x$ and $y$ components of $\vek J= \vek L + \vek S_1 + \vek S_2$ should vanish at that epoch.
In contrast, it is usual to specify freely the two spins in a Cartesian coordinate system 
where $\vek l$ points along the $z$-axis
while computing the time-domain GW polarization states for inspiraling spinning compact binaries.
Therefore, the following steps are required to evaluate the expressions for 
 $ h_{+}|_{\rm Q}(t)$ and $ h_{\times}|_{\rm Q}(t)$, given by equations~(\ref{eq:hplus_hcross}), 
specified in the $\vek j_0$-based source frame.
In the first step, the three Cartesian components of the total angular momentum at the 
initial epoch are computed and the two angles, $(\theta_j, \phi_j)$,
that specify the orientation of $\vek j_0$ in the $\vek l$-based Cartesian coordinate system are
estimated. The second step involves rotating the $\vek j_0$, $\vek l, \vek s_1$ and $\vek s_2$ vectors,
specified in the above $\vek l$ frame, by the two angles $-\theta_j$ and $ -\phi_j$.
This results in a new Cartesian coordinate system 
where $\vek j_0$ points along the $z$-axis and $\vek l$ is given by $(\sin \theta_j, 0, \cos \theta_j)$.
This is the frame where one obtains the temporally evolving $ h_{+}|_{\rm Q}(t)$ and  $ h_{\times}|_{\rm Q}(t)$
by simultaneously solving the Cartesian components of $\vek l, \vek s_1$ and $\vek s_2$ 
along with the PN-accurate differential equations for $\Phi'$ and $x$.
Further, it is easy to note that the initial estimates for $\theta_j$ and $\iota$ are equal from the definitions
of these two angles.

 We observe that the resultant $\vek j_0$ source frame is different from our source frame, depicted in figure~\ref{figure:frame}.
This is because the $x$ and $y$ axes of these two $\vek j_0$-based source frames do not usually coincide.
However, it is possible to align the $x$ and $y$ axes of these two frames by rotating our source frame 
about $\vek j_0$ such that $\vek k$  lies in the $x$-$z$ plane.
We have verified that the accumulated changes in $\Phi, \alpha$ and $\iota$ 
in the aLIGO frequency window are similar while 
evolving spinning compact binaries in these two source frames.
Finally, we note that  the two spins should not be freely specified
in the $\vek l$-based orbital triad as displayed in figure~4 of \cite{BCV}.
This is because of the equations~(38) in ~\cite{BCV} that are required to define the 
unit vectors of the 
$(\vek e_1, \vek e_2, \vek e_3 \equiv \vek l)$ triad.
It is easy to infer that 
the definition of $\vek e_1 \propto \vek l \times \vek j_0$ implies that 
$\vek j_0 \cdot \vek e_1 \equiv 0$. However, 
this dot product will not vanish  if one freely specifies the two spins 
in the above orbital triad and evaluate $\vek j_0 \cdot \vek e_1 $
by computing $\vek J $ at the initial epoch.
This anomalous feature is the main reason for stating that 
the two spins should not be freely specified at the initial epoch 
in the above orbital triad.

  A consequence of invoking $\vek k$ to specify the binary orbit is the appearance of 
new 1.5PN contributions to the amplitudes of $h_{+} $ and $h_{\times}$
in addition to what is provided by equations~(A2) and (A3) in ~\cite{ABFO}. 
These additional amplitude corrections to GW polarization states arise 
mainly due to the fact that the component of $\vek v$ along $\vek k$ is of 1.5PN order
and enter expressions for $h_{+} $ and $h_{\times}$ through the 
dot products $( \vek p \cdot \vek v )\, ( \vek q \cdot \vek v ) $ and 
$( \vek p \cdot \vek v )^2 - ( \vek q \cdot \vek v ) ^2 $ present 
in equations~(\ref{eq:h_plus_and_h_cross_in_n_and_v}).

The resulting 1.5PN order amplitude  corrections to  $h_{+} $ and $h_{\times}$ read
% \begin{widetext}
\begin{subequations}
\label{Eq_hp_hx_extra}
\begin{align}
h_{+}\Big|_{1.5\rm PN} &=\frac{G\, \mu}{c^4\, R'}\, \frac{G\, m}{2\,c^3\, \sqrt{x}}\,
 \Biggl\{  \Bigl[  \bigl(  \cos \iota 
+1 )\,(1 +C_{\theta}^2)\, \sin \iota \, \sin (2\alpha+ 2\Phi)   \nonumber \\
&\quad-(\cos \iota -1 )\, (1+C_{\theta}^2) \, \sin \iota \, \sin (2\alpha  -2\Phi)     \nonumber \\
&\quad- \big( -4\, \cos^2 \iota 
+ 2\, \cos \iota + 2 \big)\, S_{\theta} \, C_{\theta} \, \sin (\alpha -2\Phi)  \nonumber \\
&\quad+ (-2\, \cos \iota -4\, \cos^2 \iota +2)\, S_{\theta} \, C_{\theta} \sin (\alpha + 2\Phi)  \nonumber \\
&\quad-  (2+2\, C_{\theta}^2)\, \sin \iota \, \sin (2\alpha)
+4\, C_{\theta} \, S_{\theta} \, \sin \alpha\, \cos \iota    \nonumber \\
&\quad+ (-6+6\, C_{\theta}^2)\, \cos \iota \, \sin \iota\, \sin (2\Phi) 
\Bigr]\, \dot \iota  
+  \Bigl[ (-4+ 8\, \cos^2 \iota)\, \sin \iota\, S_{\theta}\, C_{\theta}\, \cos \alpha    \nonumber \\
&\quad+ (2\,  \cos \iota
+ 4 \cos^2 \iota -2)\, \sin \iota \,  S_{\theta}\, C_{\theta}\, \cos (\alpha+2\Phi)   \nonumber \\
&\quad + (\cos^3 \iota +\cos^2 \iota - \cos \iota -1)\,
 (1+C_{\theta}^2)\, \cos (2\alpha + 2\Phi)    \nonumber \\
&\quad+ (-2 +4 \cos^2 \iota -2\, \cos \iota)\, S_{\theta}\, 
C_{\theta}\, \sin \iota \,\cos (\alpha-2\Phi)     \nonumber \\ 
&\quad +(\cos^3 \iota -\cos^2 \iota -\cos \iota +1)\, 
(1+C_{\theta}^2)\, \cos (2\alpha-2\Phi) \nonumber \\
&\quad+ (-6\, \cos^2 \iota + 6)\, (1-C_{\theta}^2)\, 
\cos \iota\, \cos (2\Phi)  \nonumber \\
&\quad + (-6+ 2\, \cos (2\alpha)+2\, C_{\theta}^2\, 
\cos (2\alpha)+ 6\, C_{\theta}^2)\, \cos^3 \iota  \nonumber \\ 
&\quad + (-2\, C_{\theta}^2\, \cos(2\alpha)-6\, 
C_{\theta}^2+6-2\, \cos (2\alpha))\, \cos \iota
 \Bigr]\, \dot \alpha   
\Biggr\}   \,,                  \\ 
%%%%%%%%%%%%%%%%%%%%%%%%%%%%%%%%%%%%%%%%%%%%%%%%%%%%%%%%%%%%%%%%%%%%%%%%%%%%%%%%%%
 h_{\times}\Big|_{1.5\rm PN} &= \frac{G\, \mu}{c^4\, R'}\, \frac{G\, m}{c^3\, \sqrt{x}}\,
 \Biggl\{  \Bigl[ (2\, \cos^2 \iota  
+ \cos \iota -1)\, S_{\theta}\, \cos (\alpha + 2\Phi)   \nonumber \\
&\quad+ ( -\cos \iota - 1)\, C_{\theta} \,\sin \iota \,\cos (2\alpha+ 2\Phi) 
+( \cos \iota              
- 2\, \cos^2 \iota + 1)\, S_{\theta} \, \cos (\alpha-2\Phi) \nonumber \\
&\quad+( \cos \iota -1 )\, C_{\theta} \, \sin \iota \, \cos (2\alpha-2\Phi)  
-2\, S_{\theta} \, \cos \alpha\, \cos \iota 
+ 2 \,C_{\theta}\, \sin \iota \, \cos (2\alpha)    \Bigr]\, \dot{\iota}  \nonumber \\
&\quad+    \Bigl[ ( \cos^3 \iota\, - \cos \iota 
+ \cos^2 \iota  -1)\,C_{\theta}\, \sin (2\Phi +2\alpha) \nonumber \\
&\quad- ( \cos \iota - 2\, \cos^2 \iota 
+ 1)\, S_{\theta} \,\sin \iota\, \sin (\alpha-2\Phi)    \nonumber \\
&\quad-(- \cos^3 {\iota}+ \cos^2 \iota 
+\cos \iota - 1)\, C_{\theta} \, \sin(2\alpha-2\Phi)   \nonumber \\
&\quad+(2\,  \cos^2 \iota +  \cos \iota  
-1)\,S_{\theta}\, \sin \iota \, \sin (\alpha + 2\Phi)     
+(4\, \cos^2 \iota -2)\, S_{\theta} \, \sin \iota \sin \alpha  \nonumber \\
&\quad-2\, C_{\theta}\, \cos \iota\, \sin (2\alpha) + 2\, C_{\theta} \, \cos^3 \iota \sin (2\alpha)
  \Bigr]\, \dot{\alpha} \Biggr\}  \,,
\end{align}
\end{subequations}
% \end{widetext}
where $x$ equals the square of the invariant velocity employed in ~\cite{ABFO}.
Therefore, the fully 1.5PN order amplitude corrected $h_{+} $ and $h_{\times}$
for spinning compact binaries in quasi-circular orbits, specified by $\vek L$, are 
provided by equations~(A2) and (A3) in ~\cite{ABFO} along with 
above two 1.5PN order contributions. This requires that the $\iota$ and $\alpha$ variables
of ~\cite{ABFO} that appear in their equations~(A2) and (A3) describe $\vek k$ rather than $\vek l$.
 In contrast, the equations~(A2) and (A3) in ~\cite{ABFO} indeed
provide the fully 1.5PN order amplitude corrected $h_{+} $ and $h_{\times}$
for spinning compact binaries while invoking $\vek l$ to describe the quasi-circular orbits.
Let us state again that it will be desirable to involve equation~(\ref{Eq_lNdot}) to evolve $\vek l$ 
due to the earlier discussions.
It will be interesting to probe the influence of these  amplitude corrections 
to the parameter estimation accuracies.
We observe that 
the combined effects of spin-precession,
subdominant harmonics and amplitude modulations indeed improved
the parameter estimation accuracies of massive BH binaries that LISA should observe \cite{KJS2009}.
%http://adsabs.harvard.edu/abs/2009PhRvD..80f4027K
Additionally, our expressions for $h_{+, \times}(t)$ should be 
appealing while constructing amplitude corrected spinning binary templates invoking the 
Effective One Body (EOB) approach \cite{EOB2013}.
% http://arxiv.org/abs/1311.2544
This is because the EOB Hamiltonian approach naturally 
employs $\vek L$ to describe orbits.

  In what follows we explore, motivated purely by 
few theoretical considerations,
if it is essential to employ  
an orbital-like frequency $\omega_{\rm orb}$ while employing 
the $( {\vek n},{\vek {\xi}}, {\vek k} )$ frame to perform 
GW phasing for spinning compact binaries.  
Let us emphasize that the next section is not an alternative 
to what is detailed in this section and further investigations will 
be required to extract implications of these theoretical considerations.

\section{ Another plausible approach for GW phasing while using $\vek L $ to 
specify binary orbits }
\label{GS_way}

We begin by listing the theoretical considerations that prompted us to 
explore alternatives to an orbital-like frequency $\omega_{\rm orb}$ 
while computing inspiral templates for spinning compact binaries.
We note that the orbital-like frequency $\omega_{\rm orb}$  naturally arises in the 
$( {\vek n},{\vek {\lambda}}, {\vek l} )$ frame 
leading to the following PN independent expressions for 
binaries in circular orbits: 
$\vek v = r\, \omega_{\rm orb}\, {\vek {\lambda}}$ and 
$ \omega_{\rm orb} \equiv v/r$.  
These expressions allow us to write $ \omega_{\rm orb} \equiv \dot {\vek n} \cdot \vek \lambda $ 
which leads to 
the usually employed prescription to model 
$\Phi'(t)$, namely
$ \omega_{\rm orb} = \dot \Phi' + \cos \iota' \, \dot \alpha' $ \cite{LK_95,ABFO}.
Therefore, the phase evolution $\Phi'(t)$ becomes
\begin{equation}
\label{phiorb}
\Phi'(t) = \int^t_0 \left [\omega_{\rm orb}(t') -
\cos \iota'(t') \,\dot{\alpha'}(t')\right ]\,dt'\,,
\end{equation}
where the temporal evolution for $ \omega_{\rm orb}$ arises only due to reactive 
evolution of $x$, given by equation~(\ref{Eq_dxdt2}), while temporal evolutions for $\iota'$ and $ \alpha'$ 
are generated by the combined influences of both conservative and reactive dynamics.
However, the expression that complements $\vek v = r\, \omega_{\rm orb}\, {\vek {\lambda}}$
in our frame involves 
the conjugate momentum 
$ \sf p = \sf p_{\perp} \, \vek {\xi}$ where 
$\sf p_{\perp}^2 = \left ( \vek n \times \vek {\sf p} \right )^2 $ and therefore will not 
involve any specific orbital frequency \cite{GS11}.
The situation gets a bit more complicated due to the absence of Keplerian type parametric solution 
for the binaries in eccentric orbits \cite{GS11}. We recall that such a parametric solution 
is useful to define the orbital frequency for non-spinning compact 
binaries in both circular and eccentric orbits \cite{BDI,ABIQ}.
Finally, as noted earlier, 
%we are also influenced by the observation that 
$ \int \omega_{\rm orb}(t') dt'$ (and 
its multiples) can provide GW phase evolutions for spinning binaries 
only when orbital inclinations are tiny so that one may neglect
$\iota$ expanded amplitude corrections to $h_{+}$ and $h_{\times}$ \cite{ABFO}.
This is because in such a limit amplitude corrected GW polarization states
can be expressed in terms of $\sin \psi, \cos \psi$ and their higher 
harmonics like $\sin 2 \, \psi, \cos 2\, \psi, \cos 3\, \psi, ...$ 
where $\psi = \Phi + \alpha$.
However, in our approach 
orbital inclinations are usually not very tiny even at the initial epoch.

   Let us begin by describing, in detail, the point involving Keplerian type parametric 
solution as there exists an attempt 
to incorrectly define
$ \omega_{\rm orb}$ for spinning binaries in eccentric orbits \cite{CK10}.
Recall that it is possible to define a gauge invariant orbital frequency 
$ \omega_{\rm orb}$ for both eccentric (and circular)
 binaries with non-spinning components as noted in \cite{BDI,ABIQ}.
This  orbital frequency, defined as 
$ \omega_{\rm orb} = n (1 +k)$,  requires $n$ and $k$ 
which are  the only two
gauge invariant quantities of the orbital dynamics if expressed  in terms of 
$\epsilon=(-2\, E/\mu\, c^2)$ where $E$ is the conserved
orbital energy.
These two gauge invariant quantities are 
associated with the radial and angular part of 
Keplerian type parametric solution to the underlying conservative orbital dynamics
(the Keplerian type parametric solution exits to 3PN order 
for eccentric binaries with non-spinning components  \cite{MGS}).
The above orbital frequency allows one to have 
orbital phase varying linearly in time 
with certain additional periodic variations that vanish in the circular limit of
eccentric binaries. In other words, 
$ \phi (t) = \omega_{\rm orb} \times (t - t_0) + W(l, e_t) \equiv n \times ( 1+k) \times ( t -t_0) +  W(l, e_t)$,
where $ W(l, e_t)$ is periodic and vanishes when eccentricity $e_t$ goes to zero and $l$ is 
 defined as $l = n ( t- t_0)$
(see equations (2.6)-(2.8) in ~\cite{ABIQ}).
Therefore, it is natural to invoke $\omega_{\rm orb}$ to do GW phasing for non-spinning compact binaries
moving in eccentric orbits and to explore  numerical relativity and GW data analysis implications as
pursued in ~\cite{H10,B10}.
However, it is not possible to write down 
a similar expression for $\Phi$ in terms of $\omega_{\rm orb} $ and $W$  
while considering 
spinning compact binaries in eccentric orbits.
This is because the conservative PN dynamics of spinning compact binaries
that includes the effects of dominant order spin-orbit interactions does 
not admit a Keplerian type parametric solution in its angular sector \cite{GS11}.
This can be attributed to the non-integrable nature of the angular part of the 
associated conservative orbital dynamics.
In the absence of any naturally occurring 
orbital frequency $\omega_{\rm orb} $ in terms of $n$ and $k$, 
 \cite{GS11} invoked $\epsilon$ as the PN expansion parameter in the place of $x = ( G\, m\, \omega_{\rm orb}/c^3)^{(2/3)}$.
This forced \cite{GS11} to invoke the far-zone energy flux to directly incorporate the effects of 
radiation reaction while providing an approach to compute $h_{+, \times}(t)$ for such binaries.

  It may be useful to note that \cite{CK10} obtained an incorrect orbital frequency 
by constructing an unphysical precessing frame to accommodate
the relation $\dot \Phi' = \omega_{\rm orb} - \cos \iota' \, \dot \alpha'$,
where $ \omega_{\rm orb} = d \phi/dt$ such that the azimuthal phase $\phi$ follows 
above mentioned Keplerian type parametric 
solution.
% (see equations~(20)-(23) and (60) in ~\cite{CK10} for details). 
 It was argued in \cite{CK10} that the azimuthal phase of precessing spinning compact binaries
in eccentric orbits 
admits a Keplerian type parametric solution
in a non-inertial frame that follows the precessing orbital plane, defined by $\vek L_{\rm N}$.
This is possible if the orbital velocity in the precessing non-inertial frame
can be written as $v_{\rm prec}^2 = \dot r ^2 + r^2\, \dot \phi^2 $. A careful computation
revealed that the expression for $v_{\rm prec}^2$ is more complicated for precessing binaries
as evident from  equations~(15),(18) and (19) in \cite{GS11}. Moreover,  the appropriate equation for 
$\dot \phi$ is coupled  with the longitudinal motion,
as clear from equations~(13) in \cite{GS11}. This makes it impossible to have 
 Keplerian type 
parametric solution for the angular part of orbital dynamics for such binaries.
In the absence of any naturally occurring 
orbital frequency $\omega_{\rm orb}$, definable from a Keplerian type parametric solution, 
~\cite{GS11} employed  scaled orbital energy as their PN expansion parameter 
in the place of scaled orbital frequency.
This requires us to invoke 
the far-zone orbital energy to incorporate effects of inspiral as done in the TaylorEt approximant.
The present approach provides the circular limit of what is detailed 
in  ~\cite{GS11} while making sure that the equation for $ \dot \Phi'$ is gauge-invariant.
It may be possible to provide following reason for 
the absence of a physical orbital frequency for our binaries similar to 
what is pursued in the non-spinning binaries.
It turns out that Keplerian type parametric solution to spinning compact binaries in eccentric orbits 
exists in a precessing frame, similar to the one given by equations~(38) in ~\cite{BCV},
only in two special cases for binary dynamics under consideration \cite{KG05}.
These special cases are 
 i) binary components have equal mass and ii) only one 
component is spinning and for these two cases the $z$ component of $ \vek L$ is conserved 
as detailed in ~\cite{KG05}.
The existence of such a parametric solution allows one to define a physical orbital frequency 
$ \omega_{\rm orb} = n ( 1 + k)$ with the help of equations~(4.20) and (4.30) in ~\cite{KG05}. 
However, the $z$ component of $ \vek L$ is not 
conserved in our binaries as evident from our equation for $ \dot {\vek k }$
making the angular part non-integrable and it again leads to the  absence of a Keplerian type parametric solution.
 We would like to note that a recent computation provided a Keplerian type parametric 
solution for spinning compact binaries in eccentric orbits having their spins aligned or anti-aligned 
with $\vek L$, while incorporating higher order spin-orbit and spin-spin interactions \cite{THS13}.
% http://adsabs.harvard.edu/abs/2013CQGra..30a5007T
Additionally, another recent investigation reveals that it is possible to 
tackle the non-integrable spin-orbit dynamics with the help of certain Lie series transformation \cite{TSS2013}.
%http://adsabs.harvard.edu/abs/2013PhRvD..87f4035T
This approach obtained an 
approximative solution for the spinning binary dynamics in circular 
orbits that included the next-to-leading-order spin-orbit interactions
while expanding $\eta$ around its equal mass value, namely $\eta =0.25$.

  Further, let us clarify that it is difficult to define a frame where $ \int \omega_{\rm orb}(t') dt'$
can provide GW phase evolution for spinning compact binaries. This is demonstrated by showing that in the precessing frame 
$(\hat {\vek x}_{\rm L}, \hat {\vek y}_{\rm L}, \vek l)$ of ~\cite{ABFO}
it is not possible to have $ d \Phi'/dt= \omega$ such that the adiabatic condition for a sequence of circular orbits reads 
$ \dot {\vek n} \equiv \omega \, \vek {\lambda}$.
This is rather incompatible  with the conclusion present in the Appendix~B of ~\cite{BCV} 
as the orbital triad $( \vek {e}_1, \vek {e}_2, \vek l)$ appearing in the
above Appendix is identical to the above triad 
as evident by comparing equations~(2.9) and (2.10) of ~\cite{ABFO} with equations~(B1) of ~\cite{BCV}.
We begin by writing our equation~(\ref{eq:dndt}) for $\dot {\vek n}$ as
\begin{align}
\label{Eq_ndot_BCV}
 \dot {\vek n} &= \dot {\Phi'} \, \vek {\lambda} + \biggl \{2\frac{c^3}{G\, m}\, x^3\, \sin \Phi'\, \cot \iota'\,
 \Bigl(\delta_1 \, q\, \chi_1\, (\vek s_1 \cdot \vek n ) + \frac{\delta_2}{q}\, \chi_2\, (\vek s_2 \cdot \vek n)\Bigr)\, \vek l \biggr \} \times \vek n \,.
\end{align}
A causal comparison with equation~(B2) of ~\cite{BCV} reveals that the vectorial quantity in the curly bracket can be
identified with their $\vek {\Omega}_e $. Interestingly in our case it points along $\vek l$, provided 
we use the correct equation for  $d\vek l /dt$, namely our equation~(\ref{Eq_lNdot}). This equation
may be written as 
\begin{align}
\label{Eq_ldot_BCV}
 \dot {\vek l} &= \biggl \{2\frac{c^3}{G\, m}\, x^3\, \Bigl(\delta_1 \, q\, \chi_1\, (\vek s_1 \cdot \vek n )       
+ \frac{\delta_2}{q}\, \chi_2\, (\vek s_2 \cdot \vek n)\Bigr)\, \vek n \biggr \} \times \vek l\,,
\end{align}
and we observe that the vectorial quantity in the above curly bracket is indeed $\propto \vek n$ as required by the equations~(B4) and (B5) in ~\cite{BCV}.
However, a close inspection reveals that the curly brackets in the above two equations, namely equations~(\ref{Eq_ndot_BCV}) and (\ref{Eq_ldot_BCV}), 
are not identical and 
therefore, $\dot {\Phi'}$, appearing in equation~(\ref{Eq_ndot_BCV}), can not be equated to an orbital frequency $\omega$ required 
by the adiabatic condition : $\dot{\vek n}\equiv \omega\, \vek \lambda$.
It will be interesting to explore the consequences of these observations.
%We suspect that ~\cite{BCV} inadvertently equated  
%the vectorial quantities appearing in 
%the above two curly brackets which allowed them to state that 
%in the precessing frame $ \dot {\Phi'} \equiv \omega$.

   These observations prompted us to use $\epsilon$ 
to perform GW phasing so  that the effects of 
GW damping can also be incorporated directly with the help of PN-accurate far-zone energy flux
as done in ~\cite{GS11}.
Therefore, the present GW phasing approach
%this approach to construct temporally evolving $h_{\times,+}(t)$
%This will also make sure that the present approach to GW phasing 
may be treated as the straightforward circular limit of the prescription, detailed in  ~\cite{GS11},
to compute temporally evolving GW polarization states for 
spinning compact binaries in inspiraling eccentric orbits.
The fact that we are going to invoke $\epsilon$ to do GW phasing 
makes it similar to Taylor-Et approximant, 
introduced in ~\cite{Et} for non-spinning compact binaries.
This approximant was found to be incompatible with 
various other approximants that provided inspiral templates for non-spinning compact binaries 
and therefore  undesirable from the GW data 
analysis point of view \cite{BIPS}.
%to do GW phasing for non-spinning compact binaries 
%in ~\cite{BIPS} who also argued that it is undesirable from the GW data 
%analysis point of view. 
 However, the Taylor-Et approximant may be interesting 
while dealing with spinning compact binaries as evident 
from figure~3 in \cite{HHBG}.
%http://prd.aps.org/abstract/PRD/v78/i10/e104007
This paper compared the accumulated phase difference arising from 
numerical relativity and various PN approximants over the ten cycles before the
scaled GW frequency reached $0.1$.
The comparisons were done for equal-mass binaries whose black
holes have equal spins oriented parallel to the orbital
angular momentum.
It was noted that at the 2.5PN order (the highest PN order at which all terms are known)
the TaylorEt approximant
provided the least accumulated phase disagreements especially for high spin configurations.
%There are on-going efforts to extend this analysis for precessing and unequal mass
%binaries and to construct hybrid waveforms while invoking the 2.5PN order TaylorEt approximant.
Therefore, further investigations will be required to probe physical implications of 
our present prescription, based purely on above mentioned theoretical arguments.

  Let us first obtain the differential equation for $ \Phi$ in this approach
and it also requires  the PN-accurate relation $v/r = \dot \Phi + \dot{\alpha}\, \cos \iota $
that originates from equation~(\ref{eq:v_comonv}).
We invoke the equations~(6.1) and (6.5) in ~\cite{BBF} to obtain 
$v/r $ first in terms of $ G\,m/c^2 \,r$ and then in terms of $\epsilon $
to the 2PN order.
 We employ such a PN-accurate expression for $v/r$ to obtain the following
equation for $\dot \Phi$:
\begin{align}
\label{eq_phidot}
 \dot \Phi &=  \frac{c^3}{G\, m}\,\epsilon^{3/2}\, \Biggl\{ 1
+ \epsilon\, \Bigl(  \frac{9}{8}+ \frac{\eta}{8}\Bigr)  
-\epsilon^{3/2}\, \Bigl[ (3+X_1)\, X_1 \, \chi_1\, (\vek s_1 \cdot \vek k )  \nonumber \\
&\quad+  (3+X_2)\, X_2 \, \chi_2\, (\vek s_2 \cdot \vek k )  \Bigr]     
+\epsilon^2\, \Bigl(  \frac{11}{128}\, \eta^2 -\frac{201}{64}\, \eta +\frac{891}{128}  \Bigr)
\Biggr\} -\cos \iota \, \dot \alpha\,.
\end{align}
This is clearly equivalent to equation~(\ref{Eq_phidot}) for $\dot \Phi$ while employing PN-accurate 
expression for 
$ \omega_{\rm orb}$ in terms of $\epsilon$ as expected.

  The other aspects of 
the conservative dynamics, relevant for GW phasing, are provided by the following
precessional equations for $\vek k, \vek s_1$ and $\vek s_2$
\begin{subequations}
\label{Eq_dxy_ks1s2_xi}
\begin{eqnarray}
%\begin{align}
{\dot {\vek k}} &=&  \frac{c^3}{Gm}\, \epsilon^3\,\Bigg\{ \delta_1\,q\, \chi_1\, 
\left( \vek s_1\times\vek k\right) \label{eq:kdot_xi} 
+\frac{\delta_2}{q}\, \chi_2\, \left( \vek s_2\times\vek k\right) \Bigg\} \,, \\
{\dot {\vek s}_{1}} &=&  \frac{c^3}{Gm}\, \epsilon^{5/2}\,\delta_1 \left(\vek k\times \vek s_1\right) \,, \\
{\dot {\vek s}_{2}} &=&  \frac{c^3}{Gm}\, \epsilon^{5/2}\,\delta_2 \left(\vek k\times \vek s_2\right) \,.
%\end{align}
\end{eqnarray}
\end{subequations}
These equations are clearly obtained from their $x$ counterparts by simply substituting 
$x$ by $\epsilon$ as at the Newtonian order $x= \epsilon$.
Clearly, the above equations along with equation~(\ref{eq_phidot}) provide 
how the Eulerian angles $ \Phi, \alpha$ and $\iota$, appearing in the 
expressions for $h_{+,\times}(t)$, vary under the precessional and 
conservative dynamics in the $\epsilon-$approach.
The effect of gravitational radiation reaction is incorporated 
by allowing  $\epsilon$ to vary according to
\begin{align}
\label{eq_xidot}
 \frac{ d \epsilon}{dt} &=  \frac{64}{5}\, \frac{c^3}{G\, m}\, \eta\, \epsilon^5
\Biggl\{ 1+ \epsilon \, \Bigl( \frac{13}{336}-\frac{5}{2}\, \eta  \Bigr) 
+ 4\, \pi\, \epsilon^{3/2} \nonumber \\
&\quad + \frac{\epsilon^{3/2}}{12}\, 
\Big[ (-328\, X_1 + 135\, \sqrt{1-4\, \eta})\, X_1\, \chi_1\, (\vek s_1 \cdot \vek k) \nonumber \\
&\quad-(328\, X_2+ 135\, \sqrt{1-4\, \eta})\,X_2\, \chi_2\, (\vek s_2 \cdot \vek k)  \Big]  \nonumber \\
&\quad+\epsilon^2\, \Bigl(\frac{5}{2}\, \eta^2-\frac{12017}{2016}+\frac{117857}{18144}\Bigr)
\Biggr\} \,.
\end{align}
The above expression may be extracted from the 2PN accurate expressions for 
the far-zone energy flux and the conserved 
energy expressed in terms of $x$, given by equations~(6.4), (6.6) and (7.11) in \cite{BBF}.
It should be obvious that we need to solve together  
equations~(\ref{eq_phidot}), (\ref{Eq_dxy_ks1s2_xi}) and 
(\ref{eq_xidot}) numerically to obtain how $\Phi, \iota, \alpha$ and $\dot \Phi$
temporally vary in the present approach, while employing Cartesian components of equations~(\ref{Eq_dxy_ks1s2_xi}).
In what follows we sketch how we specify various required initial values.

 \begin{figure}
\end{figure}

\begin{figure}[!h]
\begin{center}
\includegraphics[width=100mm,height=80mm]{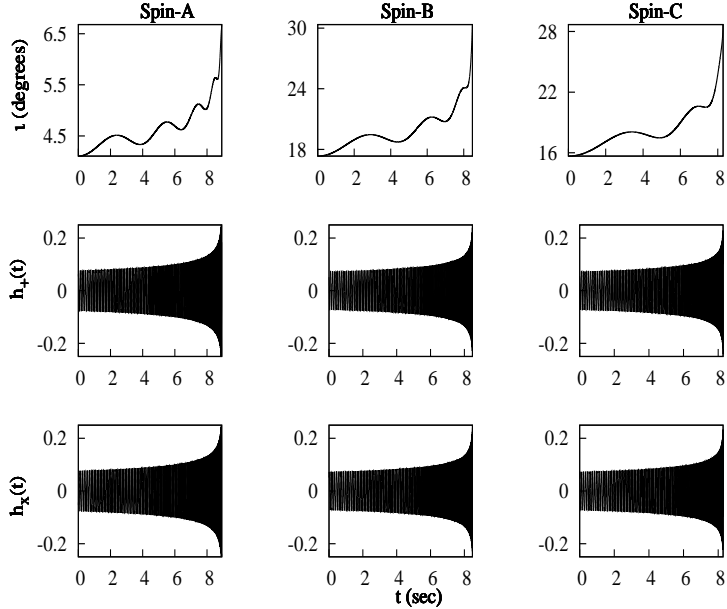}
\end{center}
\caption{Plots, similar to figure~\ref{figure:q_1_x},
displaying temporally evolving  
$\iota(t)$, scaled  $h_{+}\big|_{\text{Q}}(t)$ and 
$h_{\times}\big|_{\text{Q}}(t)$ for maximally spinning compact binaries
$(q=1, m = 50M_{\odot})$ while using $\epsilon$ as the PN
expansion parameter. 
A visual comparison with figure~\ref{figure:q_1_x} fails to provide any 
noticeable differences. 
 }
\label{figure:q_1_xi}
\end{figure}

\begin{figure}[!h]
\begin{center}
\includegraphics[width=100mm,height=80mm]{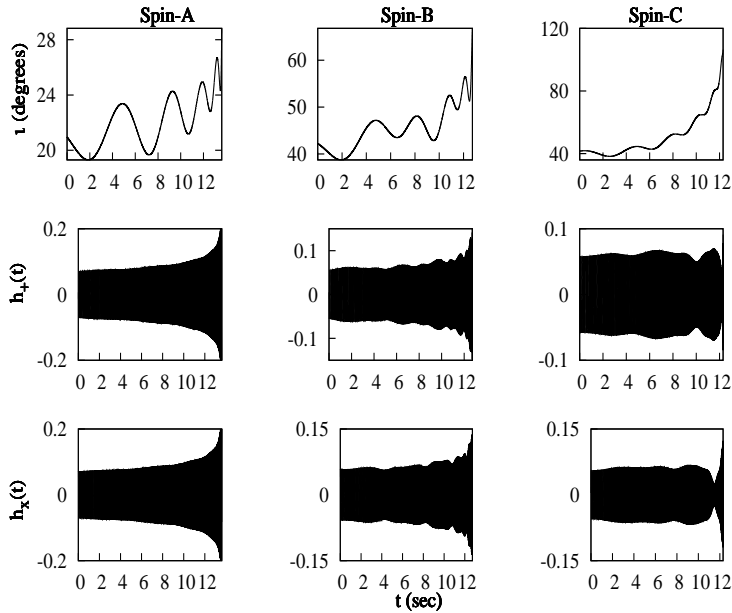}
\end{center}
\caption{ A set of plots, similar to those displayed in figure~\ref{figure:q_4_x},
 while using $\epsilon$ as the PN
expansion parameter for unequal mass maximally spinning compact binaries $(q=4, m = 50M_{\odot})$. }
\label{figure:q_4_xi}
\end{figure}

  The initial values for the Cartesian components 
of $\vek s_1$ and $\vek s_2$ in the inertial frame are provided with 
the help of equations~(\ref{eq:s1_s2}) by choosing
various possible values for $ \theta_1, \phi_1, \theta_2 $ and $\phi_2$
as expected. 
The initial values for $k_{\rm x}$ and $k_{\rm y}$ are also obtained with the help of
equations~(\ref{Eq_ialpha_ini})
% while using $ x = \e $ 
and we let $\Phi=0$ at the initial epoch.
 The bounding values for the numerical integration, namely 
the initial and
final values of $\epsilon$, are obtained by numerically inverting the following 
2PN accurate expression for $x$ in terms of $\epsilon$
\begin{align}
\label{Eq_x_2PN}
x&= \epsilon\, \Big\{ 1+\epsilon\, \Bigl( \frac{\eta}{12}
+\frac{3}{4}\Bigr) -\epsilon^{3/2}\, \Big[ \Bigl(2-\frac{2}{3}\, X_1\Bigr)
\, X_1\, \chi_1\, (\vek s_1 \cdot \vek k)        \nonumber \\
&\quad+ \Bigl(2-\frac{2}{3}\, X_2\Bigr)\, X_2\, \chi_2\, (\vek s_2 \cdot \vek k )\Big]  
+\epsilon^2\, \Bigl( \frac{\eta^2}{18}-\frac{17}{8}\, \eta+\frac{9}{2}\Bigr)    
\Big\} \,.
\end{align}
Following the previous section,
the lower limit for $\epsilon$ corresponds to $ x = 2.9\times 10^{-4}\,(m\, \omega_0)^{2/3} $ 
while the terminating value of $\epsilon$ is  obtained 
by numerically evaluating the above equation for $x$ at every epoch 
and making sure that $x$ never crosses the usual final value 
of 1/6.  
We are now in a position to obtain plots, similar to 
figures~\ref{figure:q_1_x} and \ref{figure:q_4_x}, depicting 
temporal evolutions of $\iota(t)$ and scaled $h_{+, \times}\big|_{\text{Q}}(t)$ 
while using $\epsilon$ as the PN expansion parameter and these are 
displayed in figures~\ref{figure:q_1_xi} and \ref{figure:q_4_xi}.
% A  visual comparison reveals no significant differences
%  between these two sets of figures. 

\begin{figure}[!h]
\begin{center}
\includegraphics[width=100mm,height=80mm]{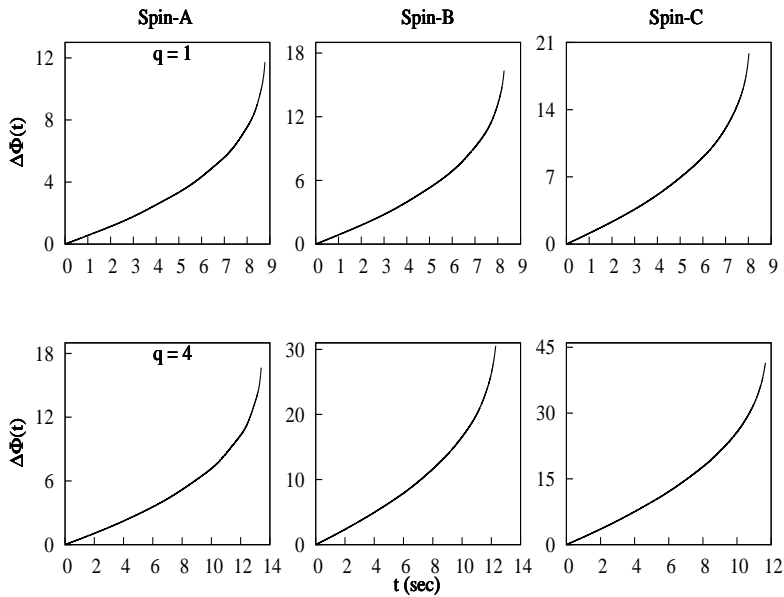}
\end{center}
\caption{The differences in the orbital phase evolution under the $x$ and $\epsilon$ approaches.
We consider two sets of binaries having $q=1$ (upper panel) and $q=4$ (lower panel) for the 
earlier discussed three initial spin configurations:
Spin-A (left panel), Spin-B (middle panel) and Spin-C (right panel).}
% and the mass ratio: $q=1$ (upper panel) and $q=4$ (lower panel).
% The $x-$approach ($\epsilon-$approach) is depicted by solid (dashed) lines. }
\label{figure:phase_evo}
\end{figure}

  A visual comparison between the plots in figure \ref{figure:q_1_x} and \ref{figure:q_1_xi} for $q=1$ binaries 
 and relevant plots in figure \ref{figure:q_4_x} and \ref{figure:q_4_xi} for $q=4$ binaries reveals no 
significant differences between the two approaches.
However,  plots for the difference in $\Phi$ evolution while 
invoking our two prescriptions show noticeable differences in their 
late time phase evolutions.
The plots in figure \ref{figure:phase_evo} show that the $ \Phi$ differences can be large.
For example, we find $\Delta \Phi \sim 40$ radians for a binary
with $q=4$ in Spin-C configuration.
It will be desirable to probe the 
differences in GW phase evolutions described by Numerical 
Relativity and our PN accurate prescriptions 
for equal and unequal mass 
precessing binaries after extending the present approaches to include 
higher order precessional and reactive dynamics. 
Let us note that such a comparison already exists 
for equal mass BH binaries having their spins aligned to orbital 
angular momentum while invoking both 
$x$ and $\epsilon$ to obtain GW phase evolutions during late stages of inspiral
 \cite{HHBG}.
These discussions should also be useful to 
probe any plausible data analysis 
implications even though non-spinning version of the $\epsilon-$approach was 
found to be undesirable in ~\cite{BIPS}.

% \begin{widetext}

% \end{widetext}

\section{Discussion}
\label{Sec_dis_con}

 We presented  a prescription to compute the time-domain GW
polarization states associated with 
 spinning compact binaries inspiraling along quasi-circular orbits.
We invoked the orbital angular momentum $\vek L$ rather than its Newtonian 
version $\vek L_{\rm N}$ to describe the binary orbits.
Additionally, we freely specified the two spins in an inertial frame 
associated with the initial direction of the total angular momentum $\vek j_0$
at the initial epoch.
In this paper, the 
precessional dynamics is governed by the leading order spin-orbit coupling
while allowing both compact objects to spin and the inspiral dynamics is 
fully 2PN accurate including the dominant order spin-orbit contributions to $d x/dt$.
We also probed the consequence of employing the precessional equation 
appropriate for $\vek L$ to numerically evolve $\vek L_{\rm N}$ and showed that 
%it leads to an undesirable feature that 
the 1.5PN order component of orbital velocity $\vek v$ along $\vek L_{\rm N}$ 
will not, in general, vanish.
We pointed out that this undesirable feature 
creates anomalous terms in the phase evolution that enter the dynamics
at the relative 3PN order.
However, our GW phase evolution is identical to what is provided 
in  ~\cite{ABFO} as both these investigations incorporate only the dominant 
1.5PN accurate spin-orbit effects.
The fact that we employ $\vek L$ to describe the binary orbits 
implies that the orbital velocity can have non-vanishing components along
$\vek L$. This leads to additional 1.5PN order amplitude corrections 
to GW polarization states compared the 1.5PN accurate amplitude 
corrected expressions for $h_{+} $ and $h_{\times}$, available in 
~\cite{ABFO} that employ $\vek L_{\rm N}$ to describe the binary orbits.
We pointed out that by adding these 1.5PN order amplitude corrections
to equations~(A2) and (A3) in ~\cite{ABFO} should result in the
fully 1.5PN order amplitude corrected $h_{+} $ and $h_{\times}$
for spinning compact binaries in 
quasi-circular orbits described by $\vek L$.
In comparison,  
the equations~(A2) and (A3) in ~\cite{ABFO} 
provide the fully 1.5PN order amplitude corrected $h_{+} $ and $h_{\times}$
for spinning compact binaries in 
quasi-circular orbits characterized by $\vek  L_{\rm N} $.
% Due to the above discussions, it will be desirable to use our equation~(\ref{Eq_lNdot}) while 
% evolving $\vek  L_{\rm N} $ to obtain inspiral templates for these binaries.

   Influenced by few purely theoretical considerations, we
 provided a plausible prescription to perform GW phasing 
that requires us to employ far-zone energy flux to implement directly 
the effects of gravitational 
radiation reaction.
The theoretical considerations include an efficient 
phasing prescription for spinning compact binaries inspiraling
along PN accurate eccentric orbits ~\cite{GS11}.
Additional considerations include the fact that 
 the orbital-like frequency
$\omega_{\rm orb}$ naturally appears in the $\vek l$ based
triad, defined using $\vek l$ and $\vek n$ and that 
 $ \int \omega_{\rm orb}(t')\, dt'$ and its multiples provide
GW phase evolutions for binaries having tiny orbital inclinations.
Further investigations will be required to probe the physical implications of 
our $\epsilon$ prescription as its non-spinning version found to be 
undesirable in ~\cite{BIPS}.

  It will be interesting to compare our numerical approach
to describe dynamics of spinning compact binaries with 
the approximate solution,  provided in \cite{TSS2013}, that 
invoked similar angular variables. 
We also plan to compare GW phase evolutions under our two 
prescriptions with those obtained from accurate NR simulations \cite{171_Cata}.
% http://adsabs.harvard.edu/abs/2013arXiv1304.6077M

\ack{
We thank Guillaume Faye and Gerhard Sch\"afer  for detailed discussions and
encouragements. 
We are grateful to P.~Ajith, B.~Sathyaprakash and N. Mazumder for clarifying the 
implementation of $h_{+,\times}(t)$ in the LSC Algorithm Library Suite.
The algebraic computations, appearing in this paper, were performed
using \textsc{Maple} and \textsc{Mathematica}.
}

\end{document}